\documentclass[aps,prd,showpacs,onecolumn,preprintnumbers,notitlepage,
nofootinbib,amssymb,10pt,amsfont,singlespacing] {revtex4-1}
\usepackage{amsmath}
\usepackage{amsmath,bm}
\usepackage{times}
\usepackage{braket}
\usepackage{color,graphicx}
\usepackage{slashed}
\usepackage{CJK,upgreek,fancyhdr}
\usepackage{hyperref}
\usepackage{multirow}
\definecolor{Red}{rgb}{1.,0.,0.}
\definecolor{Blue}{rgb}{0.,0.,1.}
\definecolor{nicered}{rgb}{0.7,0.1,0.2}
\definecolor{nicegreen}{rgb}{0.1,0.4,0.2}

\hypersetup{colorlinks,citecolor=nicegreen,linkcolor=nicered,urlcolor=magenta,anchorcolor=blue}

\newcommand{\RNum}[1]{\uppercase\expandafter{\romannumeral #1\relax}}

\begin{document}
\begin{CJK*}{GB}{gbsn}
\title{\bf Radiative decays of \boldmath{$h_{c}$} to the light mesons \boldmath{$\eta^{(\prime)}$}: A perturbative QCD calculation}

\author{Chao-Jie~Fan}
\email
[Electronic address: ]
{fancj@mails.ccnu.edu.cn}

\author{Jun-Kang~He}
\email
[Corresponding author: ]
{hejk@mails.ccnu.edu.cn}

\affiliation{Institute of Particle Physics and Key Laboratory of Quark and Lepton Physics~(MOE),\\
Central China Normal University, Wuhan, Hubei 430079, P.~R.~China}



\begin{abstract}
We study the radiative decays $h_{c}\rightarrow\gamma\eta^{(\prime)}$ in the framework of perturbative QCD and evaluate analytically the one-loop integrals with the light quark masses kept. Interestingly, the branching ratios $\mathcal{B}(h_{c}\rightarrow\gamma\eta^{(\prime)})$ are insensitive to both the light quark masses and the shapes of $\eta^{(\prime)}$ distribution amplitudes. And it is noticed that the contribution of the gluonic content of $\eta^{(\prime)}$ is almost equal to that of the quark-antiquark content of $\eta^{(\prime)}$ in the radiative decays $h_{c} \rightarrow \gamma\eta^{(\prime)}$. By employing the ratio $R_{h_{c}}=\mathcal{B}(h_{c}\rightarrow\gamma\eta)/\mathcal{B}(h_{c}\rightarrow\gamma\eta^{\prime})$, we extract the mixing angle $\phi=33.8^{\circ}\pm2.5^{\circ}$, which is in clear disagreement with the Feldmann-Kroll-Stech result $\phi=39.0^{\circ}\pm1.6^{\circ}$ extracted from the ratio $R_{J/\psi}$ with nonperturbative matrix elements $\langle 0\mid G^{a}_{\mu\nu}\tilde{G}^{a,\mu\nu}\mid\eta^{(\prime)}\rangle$, but in consistent with $\phi=33.5^{\circ}\pm0.9^{\circ}$ extracted from the asymptotic limit of the $\gamma^{\ast}\gamma-\eta^{\prime}$ transition form factor and $\phi=33.9^{\circ}\pm0.6^{\circ}$ extracted from $R_{J/\psi}$ in perturbative QCD. We also briefly discuss possible reasons for the difference in the determinations of the mixing angle.
\end{abstract}


\maketitle

\section{Introduction}
\label{sec:intro}
In recent years, the radiative decays of charmonia to the light mesons $\eta^{(\prime)}$ have been revisited in various approaches, such as perturbative quantum chromodynamics (pQCD)~\cite{Ma:2002ww,Yang:2004wy,Li:2005ug,Li:2007dq,He:2019mpy} and phenomenological models~\cite{Gerard:2004gx,Zhao:2010mm,Gerard:2013gya} (see~\cite{He:2019mpy} and references therein for more details), since they are closely related to the issue of $\eta-\eta^{\prime}$ mixing, which is important ingredient for understanding many interesting phenomena related to $\eta^{(\prime)}$. In Ref.~\cite{He:2019mpy}, the processes $J/\psi\rightarrow\gamma\eta^{(\prime)}$ were investigated and the $R_{J/\psi}=\mathcal{B}(J/\psi\rightarrow\gamma\eta^{\prime})/\mathcal{B}(J/\psi\rightarrow\gamma\eta)$ was used as the main input to extract the mixing angle $\phi=33.9^{\circ}\pm0.6^{\circ}$, which is in excellent agreement with the value $\phi=33.5^{\circ}\pm0.9^{\circ}$~\cite{Escribano:2013kba} extracted from the asymptotic limit of the $\gamma^{\ast}\gamma-\eta^{\prime}$ transition form factor (TFF) at $Q^{2}\rightarrow +\infty$, but in clear disagreement with the Feldmann-Kroll-Stech (FKS) result $\phi=39.0^{\circ}\pm1.6^{\circ}$~\cite{Feldmann:1998vh} extracted from the ratio $R_{J/\psi}$ with nonperturbative matrix elements $\langle 0\mid G^{a}_{\mu\nu}\tilde{G}^{a,\mu\nu}\mid\eta^{(\prime)}\rangle$. The difference may arise from the $g^{\ast}g^{\ast}-\eta^{(\prime)}$ TFF used in the Ref.~\cite{He:2019mpy}, in like manner, the $\gamma^{\ast}\gamma-\eta^{\prime}$ TFF used in Ref.~\cite{Escribano:2013kba}. Anyhow, more investigations are needed to provide a better understanding of the $\eta-\eta^{\prime}$ mixing.

Recently, the branching ratios of $h_{c}\rightarrow\gamma\eta^{\prime}$ and $h_{c}\rightarrow\gamma\eta$ are first measured to be, respectively, $(1.52\pm0.27\pm0.29)\times 10^{-3}$ and $(4.7\pm1.5\pm1.4)\times 10^{-4}$ with a statistical significance of $8.4 \sigma$ and $4.0 \sigma$ by the BESIII Collaboration~\cite{Ablikim:2016uoc}, where $h_c$ is assigned as $P$-wave charmonium state with the quantum numbers $J^{PC}=1^{+-}$~\cite{Tanabashi:2018oca}. In the literature, there are fewer studies on the exclusive radiative decays $h_{c} \rightarrow \gamma\eta^{(\prime)}$~\cite{Zhu:2016udl,Wu:2017pep}. In Ref.~\cite{Zhu:2016udl}, Zhu and Dai investigated these channels in the nonrelativistic QCD (NRQCD), where the contributions from the color-octet state of $h_{c}$ are suppressed by a relative factor $v_{c\bar{c}}^{2}\alpha_{s}$. And they adopted the result of the Buchmuller-Tye potential model~\cite{Buchmuller:1980su,Eichten:1995ch} for the value of the nonperturbative matrix elements, which can be directly related to the derivative of the nonrelativistic wave function at the origin~\cite{Bodwin:1994jh}. While, for $\eta^{\prime}$ production, they evaluated the contributions from the gluonic content of $\eta^{\prime}$ by the TFF $F^{(g)}_{\eta^{\prime}g^{\ast}g^{\ast}}$~\cite{Ali:2000ci}. However, in fact, the TFF $F^{(g)}_{\eta^{\prime}g^{\ast}g^{\ast}}$ enters the decay amplitude of $h_{c} \rightarrow \gamma\eta^{\prime}$ from one-loop processes. It means that Zhu and Dai~\cite{Zhu:2016udl} missed the leading order contributions of the gluonic content of $\eta^{\prime}$, which come from the tree level processes. Besides the QCD approach, Wu {\it et al.} studied the two decay processes with an intermediate meson loops model~\cite{Wu:2017pep}, and their predictions were compatible with the experimental measurements~\cite{Ablikim:2016uoc}.

From a general point of view~\cite{Appelquist:1974zd,DeRujula:1974rkb,Barbieri:1975am,Novikov:1977dq,Barbieri:1979be,Mackenzie:1981sf,Korner:1982vg,Kuhn:1983yr}, the decays of heavy quarkonium can be assumed that the annihilation of the heavy quark and antiquark is a short-distance process which can be described by perturbation theory, while the nonperturbative effect of the bound state could be factorized into its Bethe-Salpeter (B-S) wave function. In this work, we investigate the radiative decays $h_{c} \rightarrow \gamma\eta^{(\prime)}$ in the framework of pQCD and take a nonrelativistic quark model with the zero-binding approximation for the initial bound state $h_{c}$. For the final light mesons $\eta^{(\prime)}$, because of the large momentum transfer, light-cone distribution amplitudes (DAs) are adopted. As mentioned above, the issue of $\eta-\eta^{\prime}$ mixing is also involved in the decays $h_{c} \rightarrow \gamma\eta^{(\prime)}$, and we address this issue along the same line as Ref.~\cite{He:2019mpy}.

In this paper, we present a detailed calculation of the radiative decays $h_{c} \rightarrow \gamma\eta^{(\prime)}$. The involved five-point and four-point one-loop integrals are decomposed into a sum of three-point integrals and then evaluated analytically with the light quark masses kept. The branching ratios $\mathcal{B}(h_{c}\rightarrow\gamma\eta^{(\prime)})$ are found to be insensitive to the light quark masses and the shapes of the light meson DAs, which is accord with the situation in the decay processes $J/\psi\rightarrow\gamma\eta^{(\prime)}$~\cite{He:2019mpy}. And our numerical results show that the contributions from the gluonic content of $\eta^{(\prime)}$ and those from quark-antiquark content of $\eta^{(\prime)}$ are comparable with each other. The possible reasons are that: Firstly, the quark-antiquark contributions are Okubo-Zweig-Iizuka (OZI) suppressed. Specifically, the quark-antiquark contributions, which come from the one-loop OZI-forbidden processes, are suppressed by the factor $\alpha_{s}$ as compared with the gluonic contributions in the decay amplitudes. Secondly, the gluonic DA of $\eta^{(\prime)}$ can mix with their quark-antiquark DA under the QCD evolution due to the $U_{A}(1)$ anomaly~\cite{Kroll:2002nt}, which makes the gluonic part of the $\eta^{(\prime)}$ wave functions be not negligible at the charmonium scale.
In addition, we should note that, in the radiative decays $J/\psi\rightarrow \gamma\eta^{(\prime)}$~\cite{He:2019mpy,Baier:1981pm,Ma:2002ww}, the gluonic contributions are strongly suppressed by the factor $m^{2}_{\eta^{(\prime)}}/M_{J/\psi}^{2}$ because of the spin structure of their amplitudes. While similar suppressions do not exist in the decays $h_{c}\rightarrow \gamma\eta^{(\prime)}$.

The paper is organized as follows. The formalism for the decay processes $h_{c}\rightarrow\gamma\eta^{(\prime)}$ is presented in section~\ref{sec:framework}. In section~\ref{sec:numerical analysis} we present our numerical results while section~\ref{sec:conclusion} is our summary. The expressions of the eleven numerators introduced in section~\ref{sec:framework} are given in the Appendix.

\section{The radiative decays $h_{c}\rightarrow\gamma\eta^{(\prime)}$ in pQCD}
\label{sec:framework}
\subsection{The contributions of the quark-antiquark content of $\eta^{(\prime)}$}
\label{subsec:QCDq}

\begin{figure}[!!htb]
\centering
\includegraphics[width=0.45\textwidth]{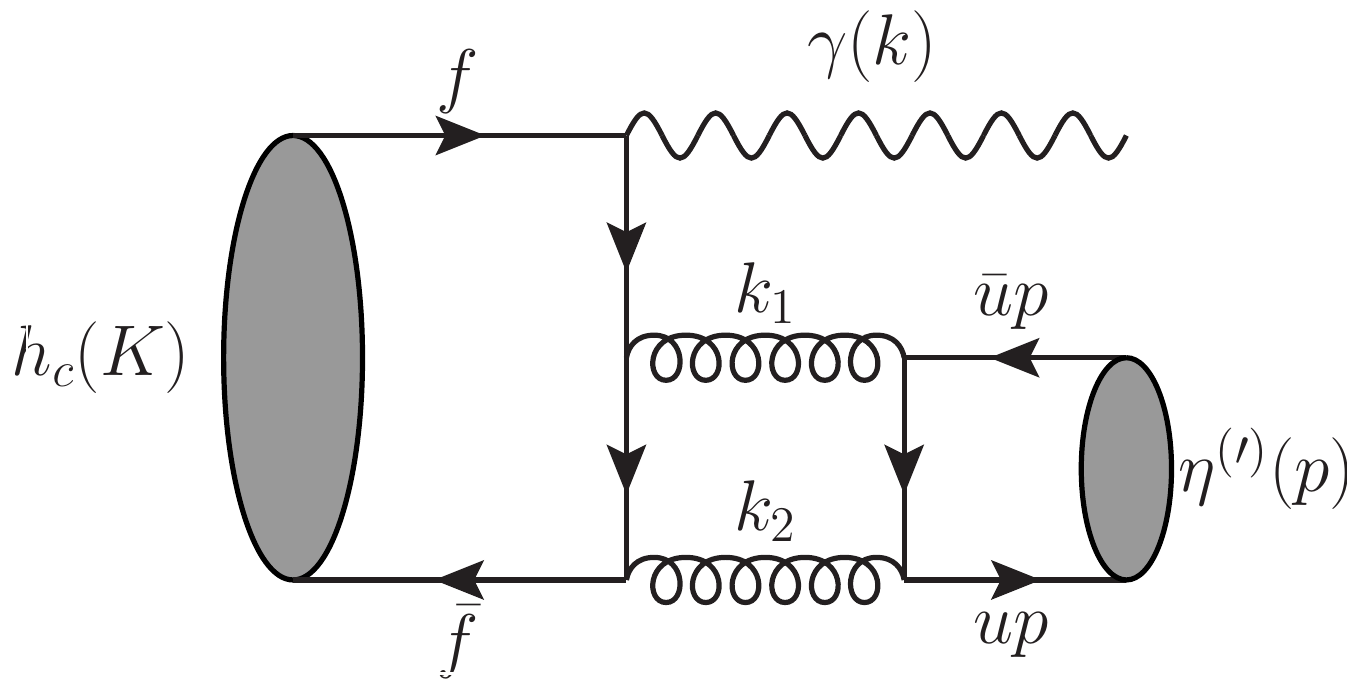}
\caption{\label{hQCDq}One typical Feynman diagram for $h_{c}\rightarrow\gamma\eta^{(\prime)}$ with the quark-antiquark content of $\eta^{(\prime)}$. Here kinematical variables are labeled.}
\end{figure}

For the quark-antiquark content of $\eta^{(\prime)}$, the leading order contributions to the radiative decays $h_{c}\rightarrow\gamma\eta^{(\prime)}$ come from one-loop processes, and one of the corresponding Feynman diagrams is depicted in Fig.~\ref{hQCDq}. There are other five diagrams from permutations of the photon and gluon legs. Following the procedure given in Ref.~\cite{Korner:1982vg}, it is convenient to divide the covariant amplitude of $h_{c}\rightarrow\gamma\eta^{(\prime)}$ into two independent amplitudes. One amplitude describes the effective coupling between $h_{c}$, a real photon and two virtual gluons, and the other describes the effective coupling between $\eta^{(\prime)}$ and two virtual gluons. Then the amplitude of $h_{c}\rightarrow\gamma\eta^{(\prime)}$ can be obtained by multiplying the two amplitudes, inserting the gluon propagators and performing the loop integration.

In the rest frame of $h_{c}$, the amplitude of $h_{c}\rightarrow \gamma g^{\ast}g^{\ast}$ can be given by~\cite{Guberina:1980xb}
\begin{eqnarray}\label{A}
A&=&A^{\alpha\beta\mu\nu}\varepsilon_{\alpha}(K)\epsilon_{\beta}^{\ast}(k)\epsilon_{\mu}^{\ast}(k_{1})\epsilon_{\nu}^{\ast}(k_{2})\nonumber\\
&=&\sqrt{3}\int\frac{\mathrm{d}^{4}q_{c}}{(2\pi)^{4}}\mathrm{Tr}\big{[}\chi(K, q_{c})\hat{\mathcal{O}}(q_{c})\big{]},
\end{eqnarray}
where $\chi(K, q_{c})$ is the B-S wave function of $h_{c}$ and $\hat{\mathcal{O}}(q_{c})$ is the hard-scattering amplitude. Here the factor $\sqrt{3}$ is included to account for the color properties of the quark and antiquark. $K$, $k$, $k_{1}$, $k_{2}$ and $\varepsilon(K)$, $\epsilon(k)$, $\epsilon(k_{1})$, $\epsilon(k_{2})$ are the momenta and polarization vectors of the $h_{c}$, the photon and the two gluons, respectively. The momenta of the quark $c$ and antiquark $\bar{c}$ have the form
\begin{eqnarray}
f=\frac{K}{2}+q_{c},\quad\quad   \bar{f}=\frac{K}{2}-q_{c}
\end{eqnarray}
with $q_{c}$ the relative momentum between the quark $c$ and antiquark $\bar{c}$.
In a nonrelativistic bound state picture, the B-S wave function $\chi(K, q_{c})$ can be reduced to its nonrelativistic form
~\cite{Kuhn:1979bb,Guberina:1980dc}
\begin{eqnarray}
\chi(K, q_{c})=2\pi\delta(q^{0}_{c})\psi_{1m}(\boldsymbol {q_{c}})\mathcal{P}(K, q_{c}),
\end{eqnarray}
where $\psi_{1m}(\boldsymbol {q_{c}})$ is the bound state wave function of $P$-wave charmonium $h_{c}$. The spin projection operator $\mathcal{P}(K, q_{c})$ can be written in a covariant form~\cite{Guberina:1980dc}
\begin{eqnarray}\label{p}
\mathcal{P}(K, q_{c})
&=&\sqrt{\frac{1}{8m_{c}^{3}}}\left(m_{c}+\frac{\slashed{K}}{2}+\slashed{q}_{c}\right)
\left(-\gamma^{5}\right)\left(m_{c}-\frac{\slashed{K}}{2}+\slashed{q}_{c}\right)
\end{eqnarray}
with $m_{c}$ the $c$ quark mass. Since the wave function $\psi_{1m}(\boldsymbol {q_{c}})$ is sharply damped for relative momenta which become large at the scale of the $h_{c}$ mass, one can expand the $\mathrm{Tr}[\mathcal{P}(K, q_{c})\hat{\mathcal{O}}(q_{c})]$ to the first order in $q_{c}$ for the $P$-wave charmonium $h_{c}$. Then the amplitude can be rewritten as
\begin{eqnarray}
A&=&\sqrt{3}\int\frac{\mathrm{d}^{3}q_{c}}{(2\pi)^{3}}\psi_{1m}(\boldsymbol{q_{c}})
q_{c}^{\alpha}\mathrm{Tr}\left[\mathcal{P}_{\alpha}(K, 0)\hat{\mathcal{O}}(0)+\mathcal{P}(K, 0)\hat{\mathcal{O}}_{\alpha}(0)\right],
\end{eqnarray}
where
\begin{eqnarray}\label{po}
\mathcal{P}_{\alpha}(K, 0)=\left.\frac{\partial}{\partial q^{\alpha}_{c}}\mathcal{P}(K, q_{c})\right|_{q_{c}=0},~~~~~~~~
\hat{\mathcal{O}}_{\alpha}(0)=\left.\frac{\partial}{\partial q^{\alpha}_{c}}\hat{\mathcal{O}}(q_{c})\right|_{q_{c}=0}.
\end{eqnarray}
With the zero-binding approximation~\cite{Kuhn:1979bb,Guberina:1980dc}, one can obtain
\begin{eqnarray}
A&=&\sqrt{3}\left(-i\sqrt{\frac{3}{4\pi}}R^{\prime}_{h_{c}}(0)\right)\varepsilon^{\alpha}(K)\mathrm{Tr}\big{[}\mathcal{P}_{\alpha}(K, 0)\hat{\mathcal{O}}(0)+\mathcal{P}(K, 0)\hat{\mathcal{O}}_{\alpha}(0)\big{]},
\end{eqnarray}
where $R^{\prime}_{h_{c}}(0)$ denotes the derivative of the radial wave function of $h_{c}$ evaluated at the origin
\begin{eqnarray}\label{zeroapp}
\int\frac{\mathrm{d}^{3}q_{c}}{(2\pi)^{3}}\psi_{1m}(\boldsymbol{q_{c}})q_{c}^{\alpha}
=-i\sqrt{\frac{3}{4\pi}}R^{\prime}_{h_{c}}(0)\varepsilon^{\alpha}(K).
\end{eqnarray}
Taking the nonrelativistic approximation $m_{c}\approx M/2$, one can obtain
\begin{eqnarray}\label{p0}
{\cal P}(K,0) &=& \sqrt{\frac{1}{4 M}} \left(\slashed{K} + M\right) \left(-\gamma^{5}\right),\nonumber\\
{\cal P}^{\alpha}(K,0) &=& \sqrt{\frac{1}{M}} \frac{\gamma^{\alpha}\slashed{K}}{M}\left(-\gamma^{5}\right),
\end{eqnarray}
and
\begin{eqnarray}\label{o0}
\hat{\mathcal{O}}(0) &=& Q_{c} \sqrt{4 \pi \alpha}\left(4 \pi \alpha_{s}\right) \frac{\delta_{ab}}{6} \frac{i}{4} \nonumber\\
& &\times\Bigg{[}\slashed{\epsilon}^{\ast}(k_{2}) \frac{\slashed{k}_{2} - \slashed{k}_{1} - \slashed{k} + M}{(k_{1}+k) \cdot k_{2}} \slashed{\epsilon}^{\ast}(k) \frac{\slashed{k}_{2} + \slashed{k} - \slashed{k}_{1} + M}{(k_{2}+k) \cdot k_{1}} \slashed{\epsilon}^{\ast}(k_{1}) + \left(\textrm{5~permutations~of~ $k_{1}$,~$k_{2}$~and~$k$}\right)\Bigg{]},\nonumber\\
\hat{\mathcal{O}}^{\alpha}(0) &=& Q_{c} \sqrt{4 \pi \alpha}(4 \pi \alpha_{s}) \frac{\delta_{ab}}{6} \frac{i}{4} \Bigg{[} \slashed{\epsilon}^{\ast}(k_{2}) \frac{2\gamma^{\alpha}}{(k_{1}+ k )\cdot k_{2}} \slashed{\epsilon}^{\ast}(k)
\frac{\slashed{k}_{2} + \slashed{k} - \slashed{k}_{1} + M}{(k + k_{2})\cdot k_{1}} \slashed{\epsilon}^{\ast}( k_{1})\nonumber \\
&&+\slashed{\epsilon}^{\ast}(k_{2}) \frac{ (k_{2}-k-k_{1})^{\alpha} (\slashed{k}_{2} - \slashed{k} - \slashed{k}_{1} + M)}{\left[(k_{1}  + k \cdot k_{2})\cdot k_{2}\right]^{2}} \slashed{\epsilon}^{\ast}(k)
\frac{\slashed{k}_{2} + \slashed{k} - \slashed{k}_{1} + M}{(k + \cdot k_{2})\cdot k_{1} } \slashed{\epsilon}^{\ast}(k_{1}
)\nonumber \\
&&+ \slashed{\epsilon}^{\ast}(k_{2}) \frac{\slashed{k}_{2} - \slashed{k}_{1} - \slashed{k} + M}{(k_{1}+k)\cdot k_{2}} \slashed{\epsilon}^{\ast}(k) \frac{ (k_{2}+k- k_{1})^{\alpha} (\slashed{k}_{2} + \slashed{k} - \slashed{k}_{1} + M)}{[(k+k_{2})\cdot k_{1}]^{2}} \slashed{\epsilon}^{\ast}( k_{1} ) \nonumber \\
&&+ \slashed{\epsilon}^{\ast}(k_{2}) \frac{\slashed{k}_{2} - \slashed{k}_{1} - \slashed{k} + M}{(k_{1}+ k)\cdot k_{2}} \slashed{\epsilon}^{\ast}(k) \frac{2\gamma^{\alpha}}{(k+k_{2})\cdot k_{1}} \slashed{\epsilon}^{\ast}( k_{1} )+ \left(\textrm{5~permutations~of~ $k_{1}$,~$k_{2}$~and~$k$}\right)\Bigg{]}.
\end{eqnarray}

\begin{figure}[!!htb]
\centering
\includegraphics[width=0.4\textwidth]{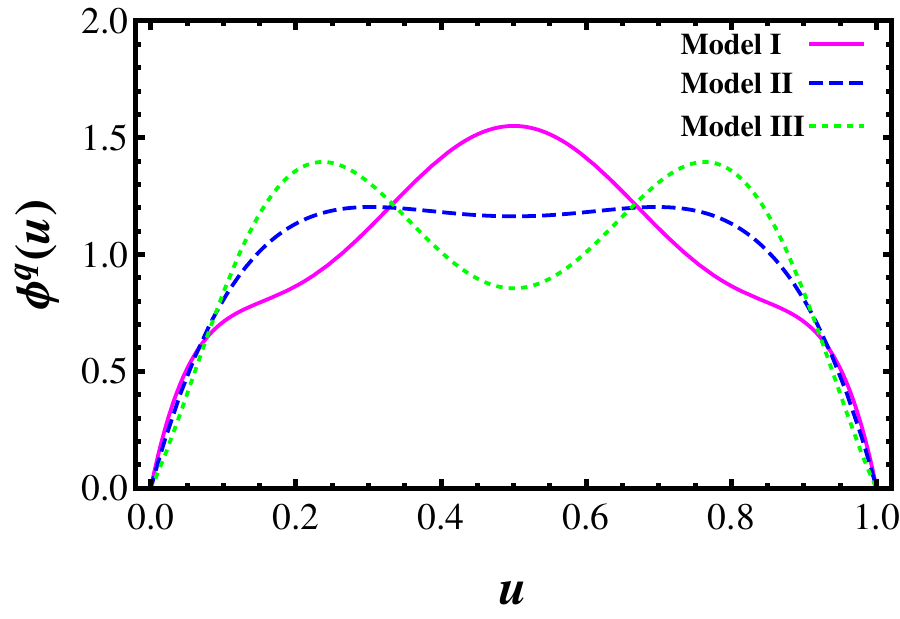}
\caption{\label{DAs}The shapes of the corresponding DAs with the scale $\mu=M/2$. }
\end{figure}

At the leading twist level, the light-cone matrix elements of the quark-antiquark content of $\eta^{(\prime)}$ are given by~\cite{Ball:2007hb}
\begin{eqnarray}
\langle\eta^{(\prime)}(p)|\bar{q}_{\alpha}(x)q_{\beta}(y)|0\rangle&=&
\frac{i}{4} f_{\eta^{(\prime)}}^{q}\left(\slashed{p}\gamma_{5}\right)_{\beta\alpha}\int\mathrm{d}u e^{i(up\cdot x+\bar{u}p\cdot y)}\phi^{q}(u), \quad\quad (q=u,d,s)
\end{eqnarray}
where the decay constants $f_{\eta^{(\prime)}}^{q}$ are defined as
\begin{eqnarray}
\langle0|\bar{q}(0)\gamma_{\mu}\gamma_{5}q(0)|\eta^{(\prime)}(p)\rangle&=&
i f_{\eta^{(\prime)}}^{q}p_{\mu}.
\end{eqnarray}
Following the conventions in Refs.~\cite{Chernyak:1983ej,Muta:1999tc,Yang:2000ce,Ali:2000ci,Yang:2004wy}, the $g^{*}g^{*}-\eta^{(\prime)}$ TFFs can be parameterized as
\begin{eqnarray}
\mathcal{M}_{\mu\nu}=-i(4\pi\alpha_{s})\delta_{ab}\epsilon_{\mu\nu\rho\sigma}
k_{1}^{\rho}k_{2}^{\sigma}F_{g^{\ast}g^{\ast}-\eta^{(\prime)}}(k_{1}^{2},k_{2}^2),
\end{eqnarray}
and the TFFs $F_{g^{\ast}g^{\ast}-\eta^{(\prime)}}(k_{1}^{2},k_{2}^2 )$ read
\begin{eqnarray}
F_{g^{\ast}g^{\ast}-\eta^{(\prime)}}(k_{1}^{2},k_{2}^2 )=\frac{1}{6}\sum_{q=u,d,s}f^{q}_{\eta^{(\prime)}}\int_{0}^{1}\mathrm{d}u
\phi^{q}(u,\mu)\left(\frac{1}{\bar{u}k_{1}^2+uk_{2}^{2}-u\bar{u}m^{2}_{\eta^{(\prime)}}
-m^{2}_{q}+i\epsilon }+(u\leftrightarrow\bar{u})\right),
\end{eqnarray}
where $\bar{u}=1-u$, $u$ is the momentum fraction carried by the quark, $m_{q}$ is the mass of the quark ($q=u,d,s$).
And the light-cone DA has the form~\cite{Agaev:2014wna}
\begin{eqnarray}
\phi^{q}(u)&=&6u(1-u)\left(1+\sum_{n=2,4\cdots}c^{q}_{n}(\mu)C_{n}^{\frac{3}{2}}(2u-1)\right)
\end{eqnarray}
with $c^{q}_{n}(\mu)$ the Gegenbauer moments.
In Table~\ref{tab:coefficients}, we list the three models of the DAs discussed in Ref.~\cite{Agaev:2014wna}. The shapes of the corresponding DAs are shown in Fig.~\ref{DAs}, where the $c^{q}_{n}(\mu)$ are evaluated at the scale $\mu=M/2$.

\begin{table}[!!htb]
  \caption{\label{tab:coefficients}Gegenbauer coefficients of three sample models with the scale $\mu_{0}=1~\mathrm{GeV}$}
\vspace{0.2cm}
\centering
  \begin{tabular}{l c r @{.} l c}
  \hline\hline
   Model    &~~~~~~~~~~~~~~$c_{2}^{q}(\mu_{0})$~~~~~~~~~~~~~~&\multicolumn{2}{c}{$c_{4}^{q}(\mu_{0})$}
            ~~~~~~~~~~~~~~&$c_{2}^{g}(\mu_{0})$    \\
  \hline
  I            &   $0.10$   &   $0$&$10$    &   $-0.26$ \\
  II           &   $0.20$   &   $0$&$00$    &   $-0.31$ \\
  III          &   $0.25$   &   $-0$&$10$   &   $-0.25$ \\
  \hline\hline
  \end{tabular}
\end{table}

As mentioned above, one can obtain the amplitude of $h_{c}\rightarrow\gamma\eta^{(\prime)}$ by multiplying the two part amplitudes, inserting the gluon propagators and integrating over the loop momentum
\begin{equation}
     M_{T}=T^{\alpha\beta}\varepsilon_{\alpha}(K)\epsilon^{\ast}_{\beta}(k)=\frac{1}{2}\int\frac{\mathrm{d}^4k_{1}}{(2 \pi)^4}A^{\alpha\beta\mu\nu}M_{\mu\nu}\varepsilon_{\alpha}(K)\epsilon^{\ast}_{\beta}(k)\frac{i}{k_{1}^2+i\epsilon}
     \frac{i}{k_{2}^2+i\epsilon},
\end{equation}
where the factor $\frac{1}{2}$ takes into account that the two gluons have already been interchanged both in $A^{\alpha\lambda\mu\nu}$ and $M_{\mu\nu}$. Using parity conservation, Lorentz invariance and gauge invariance, one can prove that
\begin{eqnarray}\label{Tmn}
T^{\alpha\beta}\sim \left(-g^{\alpha\beta}+\frac{k^{\alpha}K^{\beta}}{k\cdot K}\right),
\end{eqnarray}
i.e., there is only one independent helicity amplitude $H_{QCD}^{q}$~\cite{Korner:1982vg}
\begin{eqnarray}\label{Tee}
T^{\alpha\beta}&=&H_{QCD}^{q}h^{\alpha\beta},
\end{eqnarray}
where
\begin{eqnarray}
h^{\alpha\beta}=\left(-g^{\alpha\beta}+\frac{k^{\alpha}K^{\beta}}{k\cdot K}\right).
\end{eqnarray}
With the help of the helicity projector~\cite{Korner:1982vg}
\begin{eqnarray}
\mathbb{P}^{\alpha\beta}=\frac{1}{2}h_{\alpha^{\prime}\beta^{\prime}}\left(-g^{\alpha\alpha^{\prime}}
+\frac{K^{\alpha}K^{\alpha^{\prime}}}{M^{2}}\right)\left(-g^{\beta\beta^{\prime}}\right)=\frac{1}{2}\left(-g^{\alpha\beta}+
\frac{k^{\alpha}K^{\beta}}{k\cdot K}\right),
\end{eqnarray}
we obtain the helicity amplitude
\begin{eqnarray}\label{scalar}
H_{QCD}^{q}&=&T^{\alpha\beta}\mathbb{P}_{\alpha\beta}
=\frac{2Q_{c}}{3\sqrt{3}}\sqrt{4\pi\alpha}(4\pi\alpha_{s})^{2}\left(-i\sqrt{\frac{3}{4\pi}}
R^{\prime}_{h_{c}}\right)\sum_{q=u,d,s}\frac{f_{\eta^{(\prime)}}^{q}}{M^{\frac{5}{2}}}H_{q},
\end{eqnarray}
where the dimensionless function $H_{q}$ reads
\begin{eqnarray}\label{scahq}
H_{q}&=&\frac{1}{16\pi^{2}}\int\mathrm{d}u\phi^{q}(u)I_{q}(u).
\end{eqnarray}
$I_{q}(u)$ is the summation of the loop integrals of the six Feynman diagrams
\begin{eqnarray}
I_{q} (u) & = &\frac{1}{i \pi^{2}} \int \textrm{d}^{4} l \bigg{(} \frac{N_{1}}{D_{1} D_{2}^{2} D_{3} D_{4} D_{5}} + \frac{N_{2}}{D_{1} D_{2} D_{3}^{2} D_{4} D_{5}} + \frac{N_{3}}{D_{1} D_{2}^{2} D_{4} D_{5}} + \frac{N_{4}}{D_{1} D_{3}^{2} D_{4} D_{5}} \nonumber \\
& & + \frac{N_{5}}{D_{1} D_{2} D_{3} D_{4} D_{5}} + \frac{N_{6}}{D_{1} D_{2} D_{4} D_{5}} + \frac{N_{7}}{D_{1} D_{3} D_{4} D_{5}} \bigg{)} + (u \leftrightarrow \bar{u})
\end{eqnarray}
with $l=k_{1}-k_{2}$. Here the expressions of the denominators are given by
\begin{eqnarray}
D_{1}&=&\left(l+\xi p\right)^{2}-4m_{q}^{2}+i\epsilon,\nonumber\\
D_{2}&=&\left(l-k\right)^{2}-M^{2}+i\epsilon,\nonumber\\
D_{3}&=&\left(l+k\right)^{2}-M^{2}+i\epsilon,\nonumber\\
D_{4}&=&\left(l+p\right)^{2}+i\epsilon,\nonumber\\
D_{5}&=&\left(l-p\right)^{2}+i\epsilon,
\end{eqnarray}
and the seven numerators $N_{i}$~($i=1\thicksim7$) read
\begin{eqnarray}
N_{1} & = &  32 \left(1+x\right) M^{4}  l^{4}+ 32 M^{2} \Bigg{[}\left(3x-1\right)M^{4}-M^{2}\bigg{(}2 l\cdot k+\left(1-x\right) l\cdot p\bigg{)}\nonumber \\
& &-2l\cdot K\bigg{(}\frac{2\left(1+x \right)}{1-x} l\cdot k+l\cdot p\bigg{)}\Bigg{]}l^{2} -32 \Bigg{[}M^{6}\bigg{(}2x l\cdot k-\left(1-x\right) l\cdot p\bigg{)}\nonumber \\
& &-2M^{4}\bigg{(}3l\cdot k l\cdot p
-\frac{4 x\left(l\cdot k\right)^{2}}{1-x}-\left(l\cdot p\right)^{2} \bigg{)}-4 l\cdot k l\cdot p l\cdot K\bigg{(}M^{2}+\frac{2l\cdot K}{1-x}\bigg{)} \Bigg{]},\nonumber \\
N_{2} & = &  32 \left(1+x\right) M^{4}  l^{4}+ 32 M^{2} \Bigg{[}\left(3x-1\right)M^{4}+M^{2}\bigg{(}2 l\cdot k+\left(1-x\right) l\cdot p\bigg{)}\nonumber \\
& &-2l\cdot K\bigg{(}\frac{2\left(1+x\right)}{1-x} l\cdot k+l\cdot p\bigg{)}\Bigg{]}l^{2} +32 \Bigg{[}M^{6}\bigg{(}2x l\cdot k-\left(1-x\right) l\cdot p\bigg{)}\nonumber \\
& &+2M^{4}\bigg{(}3l\cdot k l\cdot p
-\frac{4 x\left(l\cdot k\right)^{2}}{1-x}-\left(l\cdot p\right)^{2} \bigg{)}-4 l\cdot k l\cdot p l\cdot K\bigg{(}M^{2}-\frac{2l\cdot K}{1-x}\bigg{)} \Bigg{]},\nonumber\\
N_{3} & = & 32M^{2}\Bigg{[}\left(1+x\right)M^{2}-\frac{2 x l \cdot p }{1-x}+\left(\frac{1+x}{1-x}\right)^{2}l\cdot k\Bigg{]}l^{2}-64\Bigg{(}M^{2}+\frac{2l\cdot p}{\left(1-x\right)^{2}}+\frac{l\cdot k}{1-x}\Bigg{)}l \cdot p l\cdot k\nonumber\\
& &-16\Bigg{[}
2x M^{4}l\cdot k-\left(1-x\right)M^{4}l\cdot p+4M^{2}\left(l\cdot p\right)^{2}-\frac{4\left(l\cdot p\right)^{3}}{1-x} \Bigg{]},\nonumber\\
N_{4} & = & 32M^{2}\Bigg{[}\left(1+x\right)M^{2}+\frac{2 x l \cdot p}{1-x}-\left(\frac{1+x}{1-x}\right)^{2}l\cdot k\Bigg{]}l^{2}-64\Bigg{(}M^{2}-\frac{2l\cdot p}{\left(1-x\right)^{2}}-\frac{l\cdot k}{1-x}\Bigg{)}l \cdot p l\cdot k\nonumber\\
& &+16\Bigg{[}
2x M^{4}l\cdot k-\left(1-x\right)M^{4}l\cdot p-4M^{2}\left(l\cdot p\right)^{2}-\frac{4\left(l\cdot p\right)^{3}}{1-x} \Bigg{]},\nonumber\\
N_{5} & = &  -64 M^{2}\Big{[}\left(1+x\right) M^{2} l^{2}   - 2 l \cdot p l\cdot K\Big{]} ,\quad\quad\quad N_{6}=N_{7}=  \frac{32}{1-x}\Big{[} \left(1+x\right)M^{2}l^{2} -2 l \cdot p  l \cdot K  \Big{]}
\end{eqnarray}
with $x=m^{2}/M^{2}$ and $\xi=1-2 u$.
Before going to calculate the integrals, it is useful and essential to present a short analysis of its infra-red~(IR) properties. Obviously, when one of the gluons becomes soft, e.g. $l\rightarrow p$, the propagator denominators $D_{1}$ (with $u=1-m_{q}/m$), $D_{3}$ and $D_{5}$ tend to zero, namely the individual integral may encounter soft singularities according to the conclusion in the Ref.~\cite{Dittmaier:2003bc}. However, for a given decay process, one need to go a step further and make an analysis of the numerator and denominator in the integrand simultaneously.
When $l= p+\lambda$ (the $\lambda$ is a small quantity), one can obtain the numerators
\begin{eqnarray}
N_{1}\sim N_{2}\sim N_{4}\sim \lambda^{2},\ \ \ \ N_{3}\sim N_{5}\sim N_{6}\sim N_{7}\sim \lambda,
\end{eqnarray}
and the on-shell propagator denominators
\begin{eqnarray}
D_{1}\sim D_{3}\sim \lambda,\ \ \ \ D_{5}\sim \lambda^{2}.
\end{eqnarray}
For the ultrasoft gluon ($\lambda\rightarrow 0$), the contributions to the loop function $I_{q} (u)$ have the form
\begin{eqnarray}
\int\limits_{l= p+\lambda}\textrm{d}^{4} l \bigg{(} \frac{N_{1}}{D_{1} D_{2}^{2} D_{3} D_{4} D_{5}} + \frac{N_{2}}{D_{1} D_{2} D_{3}^{2} D_{4} D_{5}} + \frac{N_{3}}{D_{1} D_{2}^{2} D_{4} D_{5}} + \frac{N_{4}}{D_{1} D_{3}^{2} D_{4} D_{5}} \nonumber \\
+ \frac{N_{5}}{D_{1} D_{2} D_{3} D_{4} D_{5}} + \frac{N_{6}}{D_{1} D_{2} D_{4} D_{5}} + \frac{N_{7}}{D_{1} D_{3} D_{4} D_{5}} \bigg{)}\sim\int\mathrm{d}^{4}\lambda\frac{\lambda}{\lambda^{4}}\rightarrow 0.
\end{eqnarray}
It means the loop function is IR safe in the potentially dangerous region.

\begin{figure}[!!htb]
\centering
\includegraphics[width=0.43\textwidth]{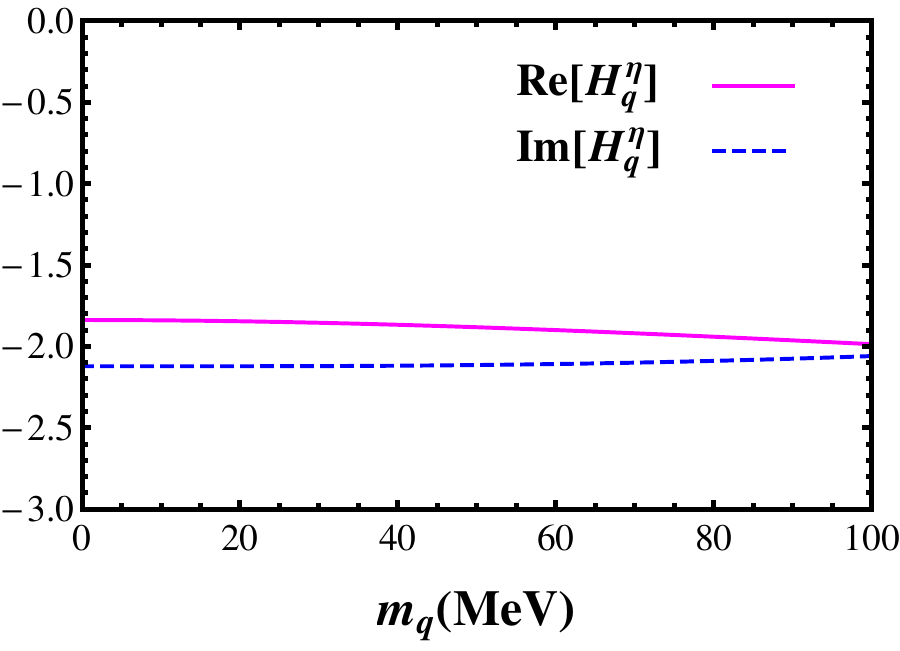}\hspace{0.5cm}
\includegraphics[width=0.43\textwidth]{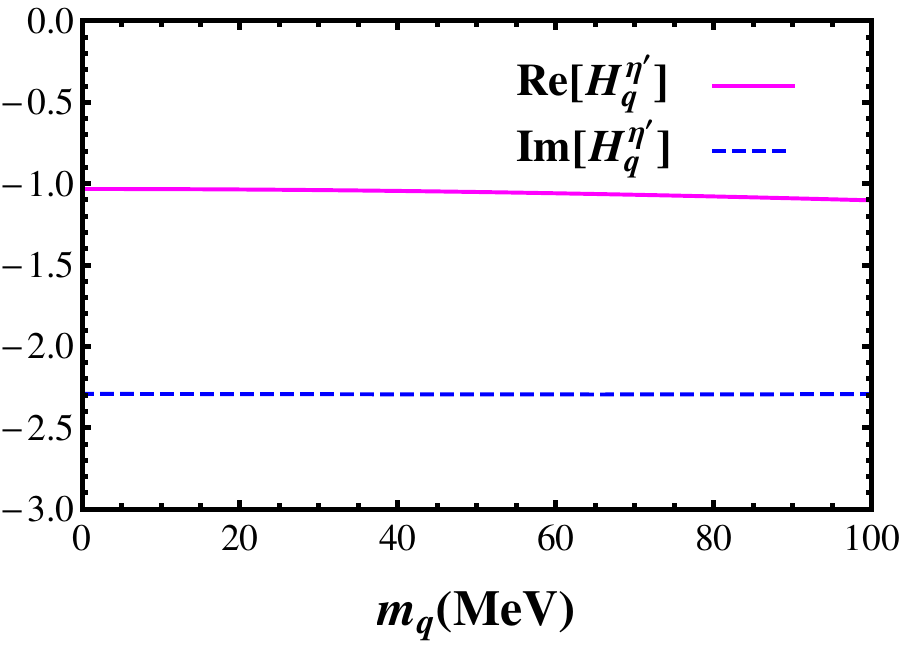}
\caption{\label{hmq}The $m_{q}$-dependence of real and imaginary parts of the dimensionless functions $H_{q}^{\eta^{\prime}}$ and $H_{q}^{\eta}$.}
\end{figure}

By using the algebraic identities
\begin{eqnarray}
l\cdot k  &=& \frac{D_{3}-D_{2}}{4},~~~~~~~~~~~~~l\cdot p=\frac{D_{4}-D_{5}}{4},~~~~~~~~~~~~~l\cdot K=\frac{D_{4}-D_{5}+D_{3}-D_{2}}{4}
\end{eqnarray}
and
\begin{align}\label{idens}
  M^{2}&=\frac{D_{4}+D_{5}}{2 \left(1+x\right)}-\frac{D_{2}+D_{3}}{2 \left(1+x\right)},&   M^{2}&=-\frac{D_{1}}{\left(1-\xi^{2}\right) x+y}
  +\frac{\left(1+\xi\right)D_{4}}{2 \left[\left(1-\xi^{2}\right)x+y\right]}+\frac{\left(1-\xi\right)D_{5}}{2 \left[\left(1-\xi^{2}\right) x+y\right]}, \nonumber\\
    l^{2}&=\frac{D_{4}+D_{5}}{2 \left(1+x\right)}+\frac{x\left(D_{2}+D_{3}\right)}{2 \left(1+x\right)},&   l^{2}&=\frac{D_{1}}{1-\xi^{2}+\frac{y}{x}} -\frac{\left[\xi \left(1+\xi\right)-\frac{y}{x}\right]D_{4}}{2\left(1-\xi^{2}+\frac{y}{x}\right)}
    +\frac{\left[\xi\left(1-\xi\right)+\frac{y}{x}\right]D_{5}}{2\left(1-\xi^{2}+\frac{y}{x}\right)}
\end{align}
with $y=4m_{q}^{2}/M^{2}$, the loop function $I_{q}(u)$ is decomposed into the sum of three-point one-loop integrals, which can be analytically calculated with the technique proposed in Ref.~\cite{tHooft:1978jhc} or the computer program $Package-\mathrm{X}$~\cite{Patel:2015tea,Patel:2016fam}. Performing the convolution integral between the loop function $I_{q}(u)$ and the DA $\phi^{q}(u)$,
our results show that the dimensionless function $H_{q}$ is insensitive to the light quark mass $m_{q}$. Specifically,
the change of the absolute value of the dimensionless function $H_{q}$ does not exceed $2\%$ when the value of $m_{q}$
varies in the range $0\sim 100\,\mathrm{MeV}$ for all the three kinds of $\eta^{(\prime)}$ DAs in Fig.~\ref{DAs}. As a consequence, the light quark mass can be neglected safely and reasonably. As shown schematically in Fig.~\ref{hmq}, we present the light quark mass $m_{q}$-dependence of the dimensionless
functions $H_{q}^{\eta}=H_{q}{\mid}_{m=m_{\eta}}$ and $H_{q}^{\eta^{\prime}}=H_{q}{\mid}_{m=m_{\eta^{\prime}}}$ with
the ``narrow" DA~(Model I in Fig.~\ref{DAs}). The property of the dimensionless function $H_{q}$ is similar with that in the radiative
decays $J/\psi\rightarrow \gamma \eta^{(\prime)}$~\cite{He:2019mpy}.

For showing the analytical expression of the loop function more clearly, we define
\begin{eqnarray}\label{scah0}
I_{0}(u)&=&\lim_{m_{q}\rightarrow 0}I_{q}(u)\nonumber\\
H_{0}&=&\frac{1}{16\pi^{2}}\int\mathrm{d}u\phi^{q}(u)I_{0}(u),
\end{eqnarray}
and then the helicity amplitude $H_{QCD}^{q}$ in Eq.~(\ref{scalar}) can be simplified to
\begin{eqnarray}\label{scalarh2}
H_{QCD}^{q}=\frac{2 Q_{c}}{3\sqrt{3}}\sqrt{4\pi\alpha}(4\pi\alpha_{s})^{2}\left(-i\sqrt{\frac{3}{4\pi}}
R^{\prime}_{h_{c}}(0)\right)\frac{f_{\eta^{(\prime)}}}{M^{\frac{5}{2}}}H_{0}
\end{eqnarray}
with the effective decay constants
\begin{eqnarray}
f_{\eta^{\prime}}=f_{\eta^{\prime}}^{u}+f_{\eta^{\prime}}^{d}+f_{\eta^{\prime}}^{s},\quad\quad
f_{\eta}=f_{\eta}^{u}+f_{\eta}^{d}+f_{\eta}^{s}.
\end{eqnarray}

\begin{figure}[!!htb]
\centering
\includegraphics[width=0.42\textwidth]{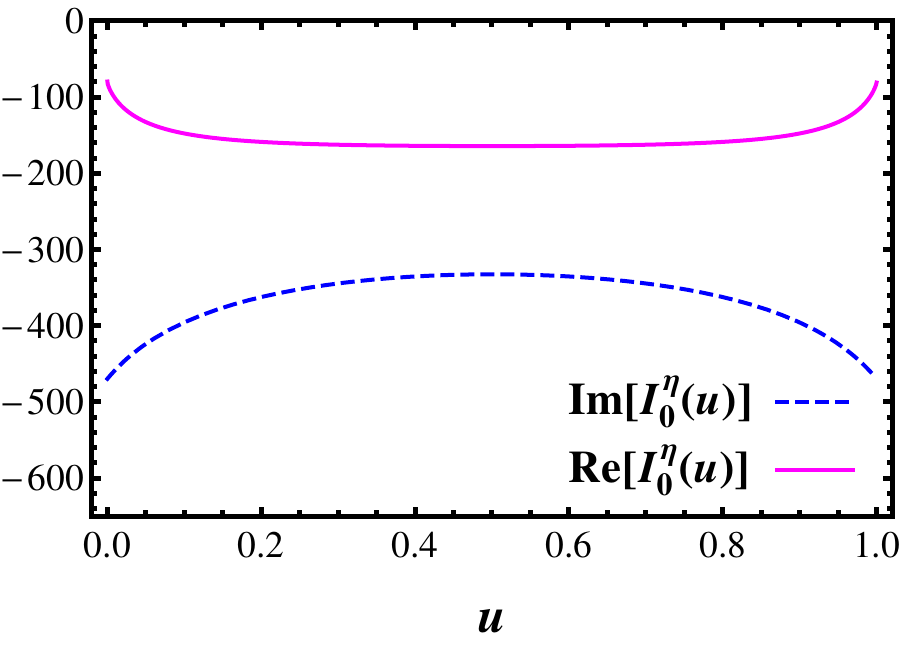}\hspace{0.5cm}
\includegraphics[width=0.42\textwidth]{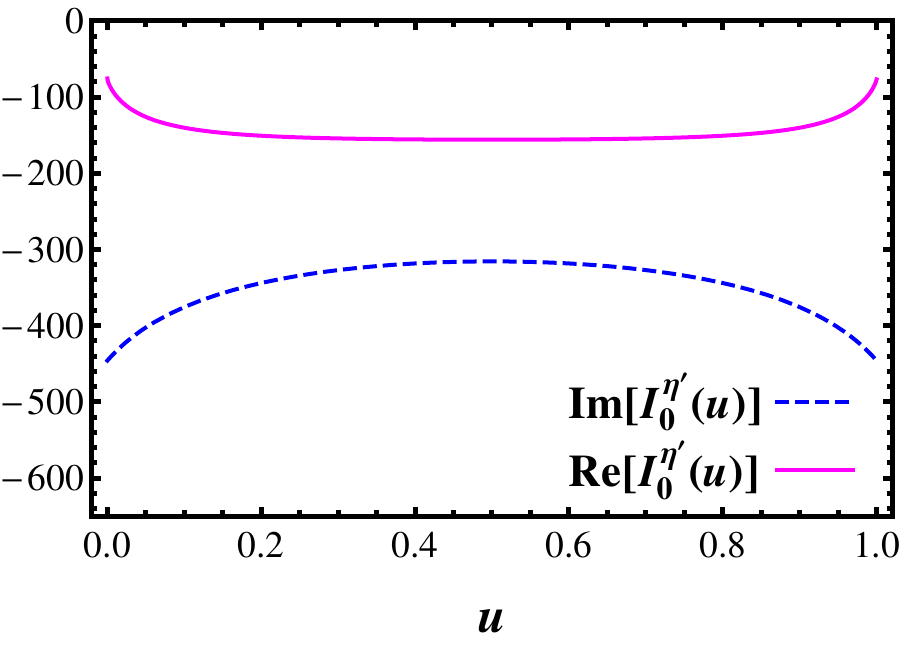}
\caption{\label{int0}The $u$-dependence of real and imaginary parts of the loop functions $I^{\eta}_{0}(u)$ and $I^{\eta^{\prime}}_{0}(u)$.}
\end{figure}

Finally, we present a brief analysis about the loop function $I_{0}(u)$. Taking the substitution $l\rightarrow -l$,
 one can obtain
\begin{eqnarray}
D_{2}\leftrightarrow D_{3},\quad\quad  D_{4}\leftrightarrow D_{5},
\end{eqnarray}
then the loop function $I_{0}(u)$ can be reduced to eleven one-loop integrals
\begin{eqnarray}
I_{0}(u)=\frac{1}{i \pi^{2}} \int \textrm{d}^{4} l& &\Bigg{(} \frac{n_{1}}{D_{1} D_{2}^{2} D_{3}} + \frac{n_{2}}{D_{1} D_{2}^{2} D_{4}} + \frac{n_{3}}{D_{1} D_{2}^{2} D_{5}} + \frac{n_{4}}{D_{2}^{2} D_{3} D_{4}} + \frac{n_{5}}{D_{2}^{2} D_{3} D_{5}} + \frac{n_{6}}{D_{2}^{2} D_{4} D_{5}}  \nonumber \\
& &+ \frac{n_{7}}{D_{1} D_{2} D_{3}} + \frac{n_{8}}{D_{1} D_{2} D_{4}}
+ \frac{n_{9}}{D_{1} D_{2} D_{5}} + \frac{n_{10}}{D_{2} D_{3} D_{4}}  + \frac{n_{11}}{D_{2} D_{4} D_{5}} \Bigg{)}
+ (u \leftrightarrow \bar{u}),
\end{eqnarray}
and the expressions of the eleven numerators $n_{i}$ ($i=1\sim 11$) are given in the Appendix.
It is noticed that the loop functions $I^{\eta}_{0}(u)=I_{0}(u){\mid}_{m=m_{\eta}}$ and $I^{\eta^{\prime}}_{0}(u)=I_{0}(u){\mid}_{m=m_{\eta^{\prime}}}$ are quite steady over the most region of $u$ as shown in Fig.~\ref{int0}. Consequently, the convolution integral between the loop function $I_{0}(u)$ and the DA becomes
insensitive to the shape of the final meson DA---in other words, it almost becomes the normalization of the DA. Specifically, the change among the dimensionless function $H_{0}$ with the different models of
the DAs in Fig.~\ref{DAs} is less than $2\%$.

\subsection{The contributions of the gluonic content of $\eta^{(\prime)}$}
\label{subsec:QCDg}
\begin{figure}[!!htb]
\centering
\includegraphics[width=0.45\textwidth]{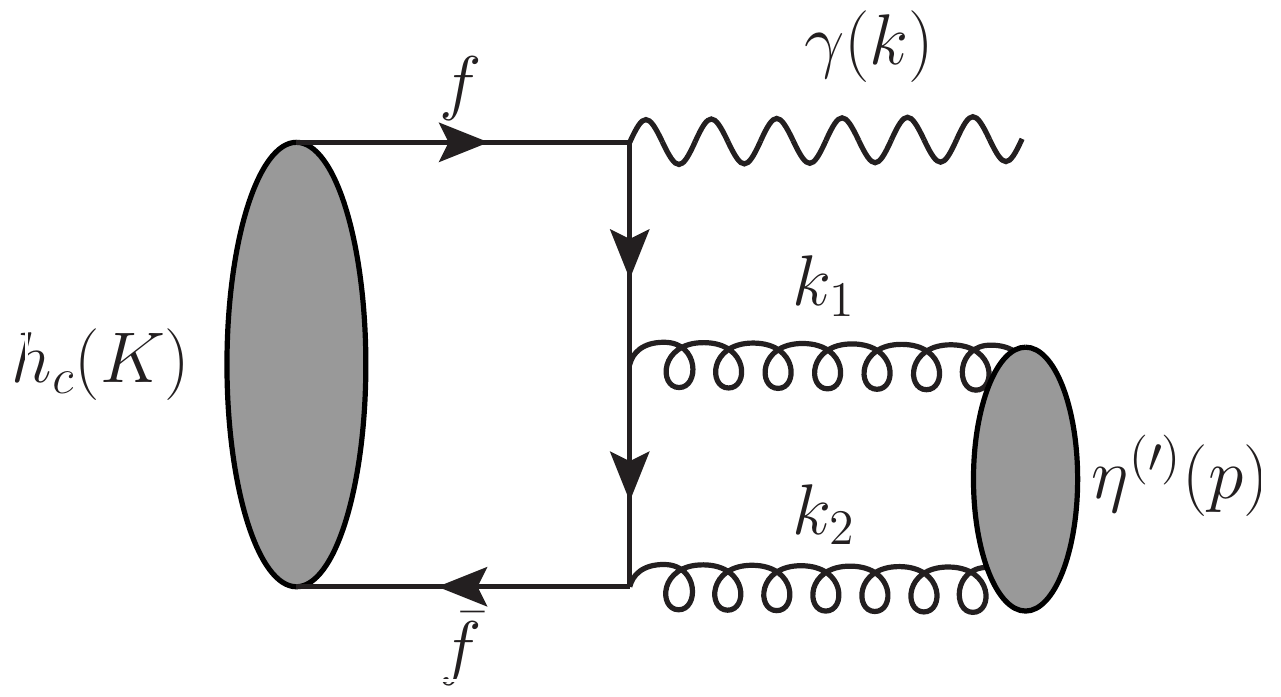}
\caption{\label{hQCDg}One typical Feynman diagram for $h_{c}\rightarrow \gamma \eta^{(\prime)}$ with the
 gluonic content of $\eta^{(\prime)}$. Here kinematical variables are labeled.}
\end{figure}

For the gluonic content of $\eta^{(\prime)}$, the leading order contributions to the radiative decays $h_{c}\rightarrow\gamma\eta^{(\prime)}$ come from the tree level processes. One of the Feynman diagrams is shown in Fig.~\ref{hQCDg},
and the other two diagrams arise from permutations of the photon and gluon legs.

The leading twist in the light-cone expansion of the matrix elements of the meson $\eta^{(\prime)}$ over two-gluon fields is~\cite{Kroll:2002nt,Ball:2007hb,Agaev:2014wna}:
\begin{eqnarray}
\langle\eta^{(\prime)}(p)|A_{\alpha}^{a}(x)A_{\beta}^{b}(y)|0\rangle&=&\frac{1}{4}\epsilon_{\alpha\beta\rho\sigma}
\frac{k^{\rho}p^{\sigma}}{p\cdot k}\frac{C_{F}}{\sqrt{3}}\frac{\delta^{ab}}{8}f_{\eta^{(\prime)}}^{1}\int\mathrm{d}u e^{i(up\cdot x+\bar{u}p\cdot y)}\frac{\phi^{g}(u)}{u(1-u)}
\end{eqnarray}
with the effective decay constant $f_{\eta^{(\prime)}}^{1}=\frac{1}{\sqrt{3}}(f_{\eta^{(\prime)}}^{u}+f_{\eta^{(\prime)}}^{d}+f_{\eta^{(\prime)}}^{s})$ and the gluonic twist-$2$ DA~\cite{Agaev:2014wna,Ball:2007hb,Alte:2015dpo}
\begin{eqnarray}
\phi^{g}(u)&=&30u^{2}(1-u)^{2}\sum_{n=2,4\cdots}c^{g}_{n}(\mu)C_{n-1}^{\frac{5}{2}}(2u-1).
\end{eqnarray}
Then we can obtain the corresponding helicity amplitude
\begin{eqnarray}
H_{QCD}^{g}&=&\frac{2Q_{c}}{9}\sqrt{4\pi\alpha}(4\pi\alpha_{s})\left(-i\sqrt{\frac{3}{4\pi}}
R^{\prime}_{h_{c}}(0)\right)\frac{f^{1}_{\eta^{(\prime)}}}{M^{\frac{5}{2}}}H_{g}
\end{eqnarray}
with
\begin{eqnarray}\label{hg}
H_{g}=\int\mathrm{d}u \frac{\phi^{g}(u)}{u(1-u)}\frac{4 (2 u-1)(2ux(1-u)-x+1)}{u(1-u)(1-x^{2}(2 u-1)^{2})}.
\end{eqnarray}

Generally, the contributions of the gluonic content are expected to be small, since the gluonic content can be seen as the higher-order effects from the point of view of the QCD evolution of the gluon DA, which vanishes in the asymptotic limit. However, due to the $U_{A}(1)$ anomaly, the gluonic DA of $\eta^{(\prime)}$ can mix with their quark-antiquark DA, which makes the contributions of the gluonic content become important in the $\eta^{(\prime)}$ production. Moreover, unlike the situation in the radiative decays $J/\psi\rightarrow\gamma\eta^{(\prime)}$~\cite{Baier:1981pm,Ma:2002ww,He:2019mpy}, in which the contributions of the gluonic content are strongly suppressed by a factor $x=m^{2}/M^{2}$, there is no additional suppression factor in the decay processes $h_{c}\rightarrow\gamma\eta^{(\prime)}$, i.e., the gluonic content of $\eta^{(\prime)}$ may play an important role in
$h_{c}\rightarrow\gamma\eta^{(\prime)}$.

\section{Numerical results}
\label{sec:numerical analysis}
In the rest frame of $h_{c}$, the decay width of $h_{c}\rightarrow\gamma\eta^{(\prime)}$ can be given by:
\begin{eqnarray}\label{width}
  \Gamma(h_{c}\rightarrow\gamma\eta^{(\prime)})=\frac{2}{3}\frac{1-x}{16\pi M}\left|H_{QCD}^{q}+H_{QCD}^{g}\right|^{2}.
\end{eqnarray}
In the numerical calculations, we employ the data given by the Particle Data Group~\cite{Tanabashi:2018oca}: $M=3525\, \mathrm{MeV}$, $m_{\eta}=548\, \mathrm{MeV}$, $m_{\eta^{\prime}}= 958\, \mathrm{MeV}$, $\Gamma_{h_c}=(0.70\pm0.28\pm0.22)\, \mathrm{MeV}$ and the decay constant $f_{\pi}=130.2\, \mathrm{MeV}$. The strong coupling constant $\alpha_{s}(M/2)=0.32$, which is calculated through the two-loop renormalization group equation.  For the derivative of the radial wave function of $h_{c}$ evaluated at the origin $R_{h_{c}}^{\prime}(0)$, we adopt the result of the Cornell potential model~\cite{Eichten:1978tg,Eichten:1979ms,Eichten:1995ch}
\begin{eqnarray}
{\mid}R_{h_{c}}^{\prime}(0){\mid}^{2}=0.131\times10^{15} \, \mathrm{MeV}^{5}.
\end{eqnarray}

For the Gegenbauer moments $c^{q}_{2}(\mu)$, $c^{q}_{4}(\mu)$, there are still large uncertainties as depicted in Table~\ref{tab:coefficients}. Fortunately, the dimensionless function $H_{0}$ in Eq.~(\ref{scah0}) is insensitive to the shapes of the $\eta^{(\prime)}$ DAs as we have shown in section~\ref{sec:framework}. So in the following numerical calculations, we choose the Model I in Table~\ref{tab:coefficients} for the DA with the scale $\mu=M/2$.

For $\eta-\eta^{\prime}$ system, the physical states $|\eta^{(\prime)}\rangle$ are usually treated as the mixing of the flavor states $|\eta_{q}\rangle=1/\sqrt{2}|u\bar{u}+d\bar{d}\rangle$ and $|\eta_{s}\rangle=|s\bar{s}\rangle$ because of the $U_{A}(1)$ anomaly. As a manifestation of the celebrated OZI-rule, one mixing angle is included in the flavor basis, and more details could be found in Refs.~\cite{Feldmann:1998vh,Agaev:2014wna}. This is the known FKS scheme~\cite{Feldmann:1998vh,Feldmann:1998sh,Feldmann:1999uf}, in which the decay constants are parameterized as
\begin{eqnarray}
 f^{u}_{\eta} &=& f^{d}_{\eta}=\frac{f_{q}}{\sqrt{2}}\cos\phi,~~~~~~~~~~ f^{s}_{\eta}=-f_{s}\sin\phi,\nonumber\\
f^{u}_{\eta'} &=& f^{d}_{\eta'}=\frac{f_{q}}{\sqrt{2}}\sin\phi,~~~~~~~~~   f^{s}_{\eta'}=f_{s}\cos\phi.
\end{eqnarray}
Here the following definitions have been used~\cite{Feldmann:1998vh,Feldmann:1998sh,Feldmann:1999uf}
\begin{eqnarray}
\langle0|J_{\mu5}^{q}(0)|\eta_{q}(p)\rangle&=&i f_{q}p_{\mu},\quad\quad
\langle0|J_{\mu5}^{q}(0)|\eta_{s}(p)\rangle=0,\nonumber\\
\langle0|J_{\mu5}^{s}(0)|\eta_{s}(p)\rangle&=&i f_{s}p_{\mu},\quad\quad
\langle0|J_{\mu5}^{s}(0)|\eta_{q}(p)\rangle=0
\end{eqnarray}
with the currents $J_{\mu5}^{q}=1/\sqrt{2}(\bar{u}\gamma_{\mu}\gamma_{5}u+\bar{d}\gamma_{\mu}\gamma_{5}d)$ and $J_{\mu5}^{s}=\bar{s}\gamma_{\mu}\gamma_{5}s$.
The mixing angle $\phi$ and the decay constants $f_{q}$, $f_{s}$ are three phenomenological parameters, which have been determined in different methods~\cite{Feldmann:1998vh,Escribano:2005qq,Escribano:2007cd,Cao:2012nj,Escribano:2013kba}.
In Table~\ref{tab:parameters}, we take the up-to-date values from Refs.~\cite{Escribano:2005qq,Escribano:2013kba}.
\begin{table}[!!htb]
  \caption{\label{tab:parameters} The values of $\phi$, $f_{q}$ and $f_{s}$ obtained with three phenomenological models~\cite{Escribano:2005qq,Escribano:2013kba}}
\vspace{0.2cm}
\centering
  \begin{tabular}{lccc}
  \hline\hline
          ~~&~~$\phi^{\circ}$~~&~~$f_{q}/f_{\pi}$~~&~~ $f_{s}/f_{\pi}$ \\
  \hline
  LEPs~\cite{Escribano:2005qq}   ~~&~~ $40.6\pm0.9$~~&~~ $1.10\pm0.03$  ~~&~~ $1.66\pm0.06$ \\
  $\eta$TFF~\cite{Escribano:2013kba}~~&~~ $40.3\pm1.8$~~&~~ $1.06\pm0.01$  ~~&~~ $1.56\pm0.24$ \\
  $\eta^{\prime}$TFF~\cite{Escribano:2013kba}    ~~&~~ $33.5\pm0.9$~~&~~ $1.09\pm0.02$  ~~&~~ $0.96\pm0.04$ \\
  \hline\hline
  \end{tabular}
\end{table}
The parameters extracted from the low energy processes (LEPs) $V\rightarrow\eta^{(\prime)}\gamma$, $\eta^{(\prime)}\rightarrow V\gamma$ ($V=\rho,\omega,\phi$) are listed in the first line. In the second line, the parameters are extracted with rational approximations for the $\eta$ TFF $F_{\gamma^{\ast}\gamma\eta}(Q^{2}\rightarrow+\infty)$. It is noteworthy that both the parameters in the first line and those in the second line are consistent with the FKS results~\cite{Feldmann:1998vh}. While in the third line, the parameters are extracted with rational approximations for the $\eta^{\prime}$ TFF $F_{\gamma^{\ast}\gamma\eta^{\prime}}(Q^{2}\rightarrow+\infty)$, which is in accord with the BaBar measurements in the timelike region at $q^{2}=112$ GeV$^{2}$~\cite{Aubert:2006cy}.

For comparison, in Tables~\ref{tab:QCDq} and~\ref{tab:QCDg},
we present the results with only the quark-antiquark contributions and those with only the gluonic contributions, respectively. Here $\Gamma_{h_c}=0.70~\mathrm{MeV}$ is adopted. The branching ratios $\mathcal{B}(h_{c}\rightarrow\gamma\eta^{\prime})$, $\mathcal{B}(h_{c}\rightarrow\gamma\eta)$ and their ratio $R_{h_{c}}=\mathcal{B}(h_{c}\rightarrow\gamma\eta)/\mathcal{B}(h_{c}\rightarrow\gamma\eta^{\prime})$ are presented in the first, second and third lines of the Tables~\ref{tab:QCDq} and~\ref{tab:QCDg}, respectively. At first glance (Fig.~\ref{hQCDq} and Fig.~\ref{hQCDg}), it seems that the branching ratios may be dominated by the ``unsuppressed" tree level
contributions from the gluonic content of $\eta^{(\prime)}$ (at order $\alpha\alpha_{s}^{2}$), rather than the one-loop contributions from the quark-antiquark
content of $\eta^{(\prime)}$ (at order $\alpha\alpha_{s}^{4}$). However, from the two tables, one can find that the quark-antiquark content and the gluonic content of $\eta^{(\prime)}$ are of almost equal importance in the radiative decays $h_{c} \rightarrow \gamma\eta^{(\prime)}$.
\begin{table}[!!htb]
  \caption{\label{tab:QCDq}The branching ratios $\mathcal{B}(h_{c}\rightarrow\gamma\eta^{(\prime)})$ with only the contributions of the quark-antiquark content}
  \vspace{0.2cm}
  \centering
  \begin{tabular}{lcccc}
  \hline\hline
              ~~~~~~~~~&~~~~~~~~~LEPs~~~~~~~~~&~~~~~~~~~$\eta$TFF~~~~~~~~~&~~~~~~~~~$\eta^{\prime}$TFF
              ~~~~~~~~~&~~~~~~~~~Exp.~\cite{Ablikim:2016uoc}~~~~~~~~~  \\
  \hline
 $\mathcal{B}(h_{c}\rightarrow\gamma\eta^{\prime})$  & $2.50\times10^{-4}$ &$2.26\times10^{-4}$&$1.32\times10^{-4}$&$(1.52\pm0.27\pm0.29)\times10^{-3}$\\
 $\mathcal{B}(h_{c}\rightarrow\gamma\eta)$&$6.5\times10^{-7}$&$1.1\times10^{-6}$&$3.6\times10^{-5}$
 &$(4.7\pm1.5\pm1.4)\times10^{-4}$\\
 $R_{h_{c}}$               &$0.3\%$ &$0.5\%$&$27.5\%$ &$(30.7\pm11.3\pm8.7)\%$  \\
  \hline\hline
  \end{tabular}
\end{table}
\begin{table}[!!htb]
  \caption{\label{tab:QCDg}The branching ratios $\mathcal{B}(h_{c}\rightarrow\gamma\eta^{(\prime)})$ with only the contributions of the gluonic content}
  \vspace{0.2cm}
  \centering
  \begin{tabular}{lcccc}
  \hline\hline
              ~~~~~~~~~&~~~~~~~~~LEPs~~~~~~~~~&~~~~~~~~~$\eta$TFF~~~~~~~~~&~~~~~~~~~$\eta^{\prime}$TFF
              ~~~~~~~~~&~~~~~~~~~Exp.~\cite{Ablikim:2016uoc}~~~~~~~~~  \\
  \hline
 $\mathcal{B}(h_{c}\rightarrow\gamma\eta^{\prime})$  & $2.50\times10^{-4}$ &$2.26\times10^{-4}$&$1.32\times10^{-4}$&$(1.52\pm0.27\pm0.29)\times10^{-3}$\\
 $\mathcal{B}(h_{c}\rightarrow\gamma\eta)$&$5.6\times10^{-7}$&$1.0\times10^{-6}$&$3.1\times10^{-5}$
 &$(4.7\pm1.5\pm1.4)\times10^{-4}$\\
 $R_{h_{c}}$               &$0.2\%$ &$0.4\%$&$23.8\%$ &$(30.7\pm11.3\pm8.7)\%$  \\
  \hline\hline
  \end{tabular}
\end{table}

In Table~\ref{tab:total}, the results with the contributions from both the quark-antiquark content and the gluonic content are presented. Here we also adopt $\Gamma_{h_c}=0.70~\mathrm{MeV}$. From Table~\ref{tab:total}, the branching ratios
$\mathcal{B}(h_{c}\rightarrow\gamma\eta^{\prime})$ and $\mathcal{B}(h_{c}\rightarrow\gamma\eta)$ are found to be greatly enhanced, because of the constructive interference of the quark-antiquark contributions and the gluonic contributions. However, both $\mathcal{B}(h_{c}\rightarrow\gamma\eta^{\prime})$ and $\mathcal{B}(h_{c}\rightarrow\gamma\eta)$ are still smaller than their experimental values. In addition, the ratio $R_{h_{c}}$ can be comparable
with its experimental value only with the $\eta^{\prime}$TFF set of parameter values.
\begin{table}[!!htb]
  \caption{\label{tab:total}The branching ratios $\mathcal{B}(h_{c}\rightarrow\gamma\eta^{(\prime)})$ with the  contributions of both the quark-antiquark content and the gluonic content}
  \vspace{0.2cm}
  \centering
  \begin{tabular}{lcccc}
  \hline\hline
               ~~~~~~~~~&~~~~~~~~~LEPs~~~~~~~~~&~~~~~~~~~$\eta$TFF~~~~~~~~~&~~~~~~~~~$\eta^{\prime}$TFF
              ~~~~~~~~~&~~~~~~~~~Exp.~\cite{Ablikim:2016uoc}~~~~~~~~~  \\
  \hline
 $\mathcal{B}(h_{c}\rightarrow\gamma\eta^{\prime})$  & $7.06\times10^{-4}$ &$6.37\times10^{-4}$&$3.73\times10^{-4}$&$(1.52\pm0.27\pm0.29)\times10^{-3}$\\
 $\mathcal{B}(h_{c}\rightarrow\gamma\eta)$&$2.0\times10^{-6}$&$3.5\times10^{-6}$&$1.1\times10^{-4}$
 &$(4.7\pm1.5\pm1.4)\times10^{-4}$\\
 $R_{h_{c}}$               &$0.3\%$ &$0.6\%$&$30.1\%$ &$(30.7\pm11.3\pm8.7)\%$  \\
  \hline\hline
  \end{tabular}
\end{table}

Considering the large uncertainties from the experimental measurement of the decay width $\Gamma_{h_c}$~\cite{Ablikim:2012ur}, the derivative of the radial wave function at the origin $R_{h_{c}}^{\prime}(0)$~\cite{Eichten:1995ch} and the factor $\alpha_{s}^{4}(\mu)$ involved in the branching ratios, it is very hard to give precise prediction for the individual branching ratio in practice. Even so, their ratio $R_{h_{c}}$ could be predicted much more reliable, since the ratio $R_{h_{c}}$ is independent of the decay width $\Gamma_{h_c}$ and the derivative of the radial wave function at the origin $R_{h_{c}}^{\prime}(0)$. Furthermore, the dependence of the ratio $R_{h_{c}}$ on $\alpha_{s}(\mu)$ is also cut down to a large extent. So we are more interested in the ratio $R_{h_{c}}$ rather than the individual branching ratio.

\begin{figure}[!!htb]
\centering
      \includegraphics[width=0.5\textwidth]{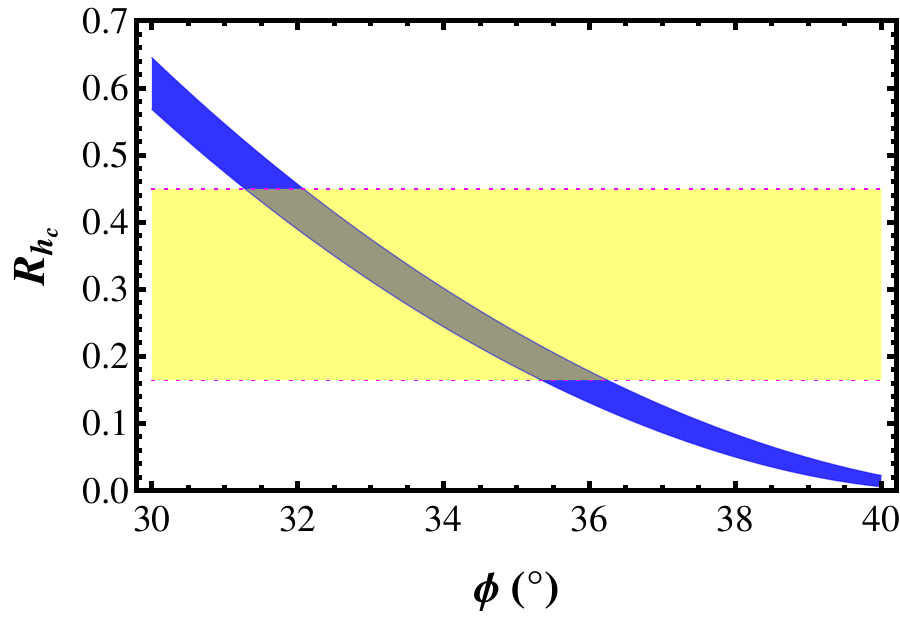}
\caption{\label{Rhc}The dependence of the ratio $R_{h_{c}}$ on the mixing angle $\phi$. The blue band is our calculated results with the uncertainties from the $\Gamma^{exp}(\eta^{(\prime)}\rightarrow\gamma\gamma)$. The yellow band denotes the experimental value of $R_{h_{c}}$ with 1$\sigma$ uncertainty.}
\end{figure}

Without inputting any phenomenological parameters, we present a determination of the mixing angle $\phi$ by the ratio $R_{h_{c}}$
\begin{eqnarray}
R_{h_{c}}=\frac{M^{2}-m_{\eta}^{2}}{M^{2}-m_{\eta^{\prime}}^{2}}
\frac{{\mid}H_{QCD}^{q}+H_{QCD}^{g}{\mid}^{2}_{m=m_{\eta}}}
{{\mid}H_{QCD}^{q}+H_{QCD}^{g}{\mid}^{2}_{m=m_{\eta^{\prime}}}}
\end{eqnarray}
and the ratio
\begin{eqnarray}
\frac{\Gamma(\eta\rightarrow\gamma\gamma)}{\Gamma(\eta^{\prime}\rightarrow\gamma\gamma)}
=\frac{m_{\eta}^{3}}{m_{\eta^{\prime}}^{3}}\left(\frac{5\sqrt{2} \frac{f_{s}}{f_{q}}-2 \tan \phi }{5\sqrt{2} \frac{f_{s}}{f_{q}} \tan \phi +2 }\right)^{2}.
\end{eqnarray}
Employing the experimental data~\cite{Ablikim:2016uoc,Babusci:2012ik,Tanabashi:2018oca}
\begin{eqnarray}
R_{h_{c}}^{exp}=(30.7\pm11.3\pm8.7)\%,\quad\quad
\Gamma^{exp}(\eta^{\prime}\rightarrow\gamma\gamma)=4.36(14)~\mathrm{KeV},\quad\quad
\Gamma^{exp}(\eta\rightarrow\gamma\gamma)=0.516(18)~\mathrm{KeV},
\end{eqnarray}
one can obtain the mixing angle and the ratio $f_{s}/f_{q}$
\begin{eqnarray}
\phi&=&33.8^{\circ}\pm2.5^{\circ},\quad\quad \frac{f_{s}}{f_{q}}=0.90\pm0.13,
\end{eqnarray}
where the big uncertainty comes mainly from $R^{exp}_{h_{c}}$~\cite{Ablikim:2016uoc}.
Schematically, we show the dependence of the ratio $R_{h_{c}}$ on the mixing angle $\phi$ in Fig.~\ref{Rhc}.

Obviously, the mixing angle $\phi$ determined by the ratio $R_{h_{c}}$ is consistent with the $\eta^{\prime}$TFF result $\phi=33.5^{\circ}\pm0.9^{\circ}$~\cite{Escribano:2013kba}, but in clear disagreement the FKS result $\phi=39.0^{\circ}\pm1.6^{\circ}$ extracted from the ratio $R_{J/\psi}$ with nonperturbative matrix elements $\langle 0\mid G^{a}_{\mu\nu}\tilde{G}^{a,\mu\nu}\mid\eta^{(\prime)}\rangle$ due to $U_{A}(1)$ anomaly dominance argument~\cite{Feldmann:1998vh} and the LEPs result $\phi=40.6^{\circ}\pm0.9^{\circ}$ extracted from the low energy processes~\cite{Escribano:2005qq}. In lattice QCD, the UKQCD collaboration~\cite{Gregory:2011sg} presented a value $\phi^{fit}\sim34^{\circ}$, while the ETM collaboration~\cite{Ottnad:2017bjt} gave $\phi=38.8^{\circ}\pm3.3^{\circ}$. The difference in the determinations of the mixing angle $\phi$ may arise from the $g^{\ast}g^{\ast}-\eta^{(\prime)}$ TFF used in our calculation, in like manner, the $\gamma^{\ast}\gamma-\eta^{\prime}$ TFF used in Ref.~\cite{Escribano:2013kba}. What is more interesting is that, in the same framework of pQCD, the mixing angles extracted from the ratio $R_{h_{c}}$ ($\phi=33.8^{\circ}\pm2.5^{\circ}$) and the ratio $R_{J/\psi}$ ($\phi=33.9^{\circ}\pm0.6^{\circ}$)~\cite{He:2019mpy} are in excellent agreement with each other. In addition, the central value of the ratio $f_{s}/f_{q}$ obtained in both this work and Refs.~\cite{Escribano:2013kba,He:2019mpy} is smaller than unity. While in the LEPs $V\rightarrow\eta^{(\prime)}\gamma$, $\eta^{(\prime)}\rightarrow V\gamma$ ($V=\rho,\omega,\phi$)~\cite{Feldmann:1998vh,Escribano:2005qq}, one usually predicted $f_{s}>f_{q}$, which is different from the result obtained in this work. However, from another point of view, the discrepancies in these determinations may indicate that our understanding of $\eta-\eta^{\prime}$ mixing scheme is incomplete, and the physical picture of the $\eta-\eta^{\prime}$ mixing at the high energy scale may differ from that at the low energy scale. Anyhow, the physics associated with the $\eta-\eta^{\prime}$ mixing is interesting, and it is certainly worthy of further investigations for a better understanding of many phenomena in $\eta^{(\prime)}$ production processes.

\section{Summary}
\label{sec:conclusion}
In this work, we have calculated the branching ratios of the radiative decays $h_{c}\rightarrow \gamma\eta^{(\prime)}$ in the framework of pQCD. For the initial heavy quarkonium $h_{c}$, we neglect its internal momentum in a nonrelativistic picture. While the final light mesons $\eta^{(\prime)}$ are treated as a light-cone object, and we employ a set of DAs of the quark-antiquark content and those of the gluonic content as nonperturbative inputs. Using some algebraic identities, we decompose the five-point and four-point one-loop integrals into the three-point one-loop integrals and calculate these three-point one-loop integrals analytically. Our results of the branching ratios $\mathcal{B}(h_{c}\rightarrow\gamma\eta^{(\prime)})$ are insensitive to the light quark masses and the shapes of the $\eta^{(\prime)}$ DAs, which is similar to the situation in the decay processes $J/\psi\rightarrow\gamma\eta^{(\prime)}$~\cite{He:2019mpy}. Furthermore, we find that the contributions from the quark-antiquark content are almost equal to those from the gluonic content, which are strongly suppressed in the heavy quarkonium decays $V\rightarrow \gamma\eta^{(\prime)}$~\cite{Baier:1981pm,Ma:2002ww,He:2019mpy}. Interestingly enough, only with the set of $\phi$, $f_{q}$ and $f_{s}$ extracted from the asymptotic limit of the $\eta^{\prime}$ TFF, which is in accord with the BaBar measurement at $q^{2}=112$ GeV$^{2}$~\cite{Aubert:2006cy}, our result of the ratio $R_{h_{c}}$ is comparable with its experimental data. As a crossing check, by using the $R_{h_{c}}$, $\Gamma(\eta^{(\prime)}\rightarrow\gamma\gamma)$ and their experimental values, we obtain the mixing angle $\phi=33.8^{\circ}\pm2.5^{\circ}$, which is in excellent consistent with the $\eta^{\prime}$TFF result $\phi=33.5^{\circ}\pm0.9^{\circ}$~\cite{Escribano:2013kba} and the result $\phi=33.9^{\circ}\pm0.6^{\circ}$ extracted by the ratio $R_{J/\psi}$~\cite{He:2019mpy}.

For the individual branching ratios $\mathcal{B}(h_{c}\rightarrow\gamma\eta^{\prime})$ and $\mathcal{B}(h_{c}\rightarrow\gamma\eta)$, there are still considerable discrepancies between our results and the experimental measurements. The reason may be due to the sensitivity of the hard-scattering amplitude $\hat{\mathcal{O}}(q_{c})$ to the internal momentum $q_{c}$ in the $P$-wave decays, which can make the convergence of the $q_{c}$-expansion of $\mathrm{Tr}[\mathcal{P}(K, q_{c})\hat{\mathcal{O}}(q_{c})]$ become poor. It is well known that there are IR divergences in the color-singlet state contributions for the inclusive $P$-wave charmonia decays~\cite{Barbieri:1976fp,Bodwin:1992ye}, and these IR divergences can be removed by considering higher-order contributions. While for the exclusive $P$-wave charmonia decays, the same IR divergences always do not appear. However, as pointed out in Refs.~\cite{Kroll:1997vt,Wong:1998rv}, the higher-order contributions, such as the higher Fock-state contributions and the relativistic corrections, are still important to the exclusive $P$-wave charmonia decays. It indicates that the nonperturbative effects beyond those contained in $R_{h_{c}}^{\prime}(0)$ may also play an important role in the radiative decays $h_{c}\rightarrow\gamma\eta^{(\prime)}$, even though the results with the zero-binding approximation for these decays are IR safe.  Within a B-S equation approach, we would revisit the decays $h_{c}\rightarrow\gamma\eta^{(\prime)}$ by retaining the relative momentum $q_{c}$ in the hard-scattering amplitude (i.e., the relativistic corrections), and this work is under way.

\bigskip
\begin{acknowledgments}
This work is supported by the National Natural Science Foundation of China under Grant Nos.~11675061, 11775092 and 11435003.
\end{acknowledgments}

\newpage
\appendix
\section*{Appendix: The expressions of the eleven numerators $n_{i}$ ($i=1\sim 11$)}
\label{sec:twenty-one numerators}
The expressions of the numerators $n_{i}$ read
\begin{align}\label{n1n11}
    n_{1}=&\frac{64 M^2 \xi^2}{\left(1-\xi^{2}\right) \left(1-x^{2} \xi^{2}\right)}\Bigg{(} \left(1-3 x\right)l^{2} +2\left(1+x\right)l\cdot k -\left(1+x\right) M^2\Bigg{)} ,\nonumber\\
    n_{2}=&-16M^2\Bigg{\{}\left(\frac{1-2 x}{x \left(1-\xi^{2}\right)}-\frac{x \left(1-\xi^{2}\right)+4\left(1-x\right)}{x \left(1-\xi^{2}\right) \left(1-x^{2} \xi^{2}\right)}\right) l^{2}-\frac{4 l\cdot p\big{[}x l\cdot k-\left(1-x\right)l\cdot p\big{]}}{\left(1-x\right)^{2}M^2} \nonumber\\
    &-\frac{2}{\left(1-x\right)^2}\left(2x^{2}-\frac{2x\left(x\xi^{2}+2\right)}{ 1-x^{2} \xi^{2}}+\frac{2x}{ 1-\xi^{2}}-\frac{3\xi^{2}\left(1-x^{4}\right)}{ \left(1-\xi^{2}\right) \left(1-x^{2} \xi^{2}\right)}\right) l\cdot k\nonumber\\
    &-2  \left(\frac{2 x \xi^2 }{\left(1-x\right) \left(1-\xi^{2}\right)}-\frac{x^{2} \xi^2 }{1-x^{2} \xi^{2}}\right) l\cdot p
    +\left(\frac{x \xi^2}{1-\xi^{2}}+\frac{x^{2} \xi^2}{1-x^{2} \xi^{2}} \right)M^2 \Bigg{\}},\nonumber\\
     n_{3}=&-16M^2\Bigg{\{}\left(\frac{1-2 x}{x \left(1-\xi^{2}\right)}-\frac{x \left(1-\xi^{2}\right)+4\left(1-x\right)}{x \left(1-\xi^{2}\right) \left(1-x^{2} \xi^{2}\right)}\right) l^{2}+\frac{4 l\cdot p\big{[}x l\cdot k-\left(1-x\right)l\cdot p\big{]}}{\left(1-x\right)^{2}M^2} \nonumber\\
    &-\frac{2}{\left(1-x\right)^2}\left(4x-1+\frac{2x-3}{ 1-\xi^{2}}-\frac{2x^{2}\left(1-x^{2}\xi^{4}\right)+x^{2}\xi^{2}\left(\xi^{2}-x^{2}\right)}{ \left(1-\xi^{2}\right) \left(1-x^{2} \xi^{2}\right)}\right) l\cdot k\nonumber\\
    &-2  \left(\frac{2 x \xi^2 }{\left(1-x \right) \left(1-\xi^{2}\right)}+\frac{x^{2} \xi^2 }{1-x^{2} \xi^{2}}\right) l\cdot p
    +\left(\frac{x \xi^2}{1-\xi^{2}}+\frac{x^{2} \xi^2}{1-x^{2} \xi^{2}} \right)M^2 \Bigg{\}},\nonumber\\
    n_{4}=&-\frac{32M^{2}\left(1+x \xi^{2}\right)}{\left(1-\xi^{2}\right) \left(1-x^{2} \xi^{2}\right)}
    \Bigg{(} \frac{1-3 x}{1+x}l^{2} +\frac{2\left(1-3x\right)}{1-x}l\cdot k-2 l\cdot p - M^2\Bigg{)}, \nonumber\\
     n_{5}=&-\frac{32M^{2}\left(1+x \xi^{2}\right)}{\left(1-\xi^{2}\right) \left(1-x^{2} \xi^{2}\right)}
    \Bigg{(} \frac{1-3 x}{1+x}l^{2} +\frac{2\left(1+x \right)}{1-x}l\cdot k+2 l\cdot p - M^2\Bigg{)}, \nonumber\\
     n_{6}=&-\frac{32M^{2}}{1-\xi^{2}}
    \Bigg{(} \frac{2x^{2}+x+3}{x\left(1+x\right)}l^{2}-\frac{6x^{2}-4x+6}{\left(1-x\right)^{2}}l\cdot k+\frac{4x}{1-x}l\cdot p - x M^2\Bigg{)}, \nonumber\\
     n_{7}=&\frac{64\xi}{\left(1-\xi^{2}\right) \left(1-x^{2} \xi^{2}\right)^{2}}
    \Bigg{(} \left(x^2 \xi^2+2 x \xi^2+1\right)l\cdot p +4 x\xi M^2\Bigg{)}, \nonumber\\
    n_{8}=&\frac{16}{1-x}\Bigg{\{}2 l\cdot k+\left(\frac{4\left(1-x \right)}{\left(1-x^{2} \xi^{2}\right)^{2}}+\frac{2x \left(1-x \xi^{2}\right)}{1-x^{2} \xi^{2}}\right)l\cdot p+\Bigg{(}\frac{10x-4\xi^{2}}{\left(1-\xi^{2}\right)\left(1-x^{2} \xi^{2}\right)}\nonumber\\
    &+\frac{\left(x^{4}\xi^{4}-3x\xi^{2}-11x+1\right)\left(1-x \xi^{2}\right)}{\left(1-\xi^{2}\right)\left(1-x^{2} \xi^{2}\right)^{2}}+\frac{x^{3}\xi^{2}\left(2x^{2}\xi^{2}+3x\xi^{2}+1\right)}{\left(1-x^{2} \xi^{2}\right)^{2}} \Bigg{)}M^{2}\Bigg{\}},\nonumber\\
     n_{9}=&-\frac{16}{1-x}\Bigg{\{}2 l\cdot k+\left(\frac{4\left(1-x \right)}{\left(1-x^{2} \xi^{2}\right)^{2}}+\frac{2x \left(1-x \xi^{2}\right)}{1-x^{2} \xi^{2}}\right)l\cdot p-\Bigg{(}\frac{x^{2}\xi^{2}\left(x^{2}\xi^{2}+2x+15\right)}{\left(1-x^{2} \xi^{2}\right)^{2}}\nonumber\\
    &-\frac{10x^{2}\xi^{2}\left(x-\xi^{2}\right)+6x\xi^{2}\left(1-x\xi^{2}\right)}{\left(1-\xi^{2}\right)\left(1-x^{2} \xi^{2}\right)^{2}}+\frac{x^{3}\xi^{4}-x^{2}\xi^{2}-2\xi^{2}+x-1}{\left(1-\xi^{2}\right)\left(1-x^{2} \xi^{2}\right)} \Bigg{)}M^{2}\Bigg{\}},\nonumber\\
    n_{10}=&-\frac{64\left(x^{2}\xi^{2}+2x\xi^{2}+1\right)}{\left(1-\xi^{2}\right)\left(1-x^{2} \xi^{2}\right)^{2}}l\cdot p
    -\frac{128x}{\left(1+x\right)^{2}}\left(\frac{2\left(1+x\right)\left(1+x\xi^{2}\right)}{\left(1-\xi^{2}\right)\left(1-x^{2} \xi^{2}\right)^{2}}-\frac{x}{1-x^{2} \xi^{2}}\right)M^{2},\nonumber\\
    n_{11}=&\frac{32 M^2 \left(x^3+x^2+11 x+3\right)}{(1-x) (1+x)^2 \left(1-\xi^2\right)}.
\end{align}

\newpage


\begin{thebibliography}{1}
\expandafter\ifx\csname natexlab\endcsname\relax\def\natexlab#1{#1}\fi
\expandafter\ifx\csname bibnamefont\endcsname\relax
  \def\bibnamefont#1{#1}\fi
\expandafter\ifx\csname bibfnamefont\endcsname\relax
  \def\bibfnamefont#1{#1}\fi
\expandafter\ifx\csname citenamefont\endcsname\relax
  \def\citenamefont#1{#1}\fi
\expandafter\ifx\csname url\endcsname\relax
  \def\url#1{\texttt{#1}}\fi
\expandafter\ifx\csname urlprefix\endcsname\relax\def\urlprefix{URL }\fi
\providecommand{\bibinfo}[2]{#2}
\providecommand{\eprint}[2][]{\url{#2}}

\bibitem[{\citenamefont{Ma}(2002)}]{Ma:2002ww}
\bibinfo{author}{\bibfnamefont{J.~P.} \bibnamefont{Ma}},
  \bibinfo{journal}{Phys. Rev.} \textbf{\bibinfo{volume}{D65}},
  \bibinfo{pages}{097506} (\bibinfo{year}{2002}), \eprint{arXiv:hep-ph/0202256}.

\bibitem[{\citenamefont{Yang}(2004)}]{Yang:2004wy}
\bibinfo{author}{\bibfnamefont{Y.-D.} \bibnamefont{Yang}}
  (\bibinfo{year}{2004}), \eprint{arXiv:hep-ph/0404018}.

\bibitem[{\citenamefont{Li et~al.}(2005)\citenamefont{Li, Li, Li, Ma, and
  Zhao}}]{Li:2005ug}
\bibinfo{author}{\bibfnamefont{G.}~\bibnamefont{Li}},
  \bibinfo{author}{\bibfnamefont{T.}~\bibnamefont{Li}},
  \bibinfo{author}{\bibfnamefont{X.-Q.} \bibnamefont{Li}},
  \bibinfo{author}{\bibfnamefont{W.-G.} \bibnamefont{Ma}}, \bibnamefont{and}
  \bibinfo{author}{\bibfnamefont{S.-M.} \bibnamefont{Zhao}},
  \bibinfo{journal}{Nucl. Phys.} \textbf{\bibinfo{volume}{B727}},
  \bibinfo{pages}{301} (\bibinfo{year}{2005}), \eprint{arXiv:hep-ph/0505158}.

\bibitem[{\citenamefont{Li}(2008)}]{Li:2007dq}
\bibinfo{author}{\bibfnamefont{B.~A.} \bibnamefont{Li}},
  \bibinfo{journal}{Phys. Rev.} \textbf{\bibinfo{volume}{D77}},
  \bibinfo{pages}{097502} (\bibinfo{year}{2008}), \eprint{arXiv:0712.4246}.

\bibitem[{\citenamefont{He and Yang}(2019)}]{He:2019mpy}
\bibinfo{author}{\bibfnamefont{J.-K.} \bibnamefont{He}} \bibnamefont{and}
  \bibinfo{author}{\bibfnamefont{Y.-D.} \bibnamefont{Yang}},
  \bibinfo{journal}{Nucl. Phys.} \textbf{\bibinfo{volume}{B943}},
  \bibinfo{pages}{114627} (\bibinfo{year}{2019}), \eprint{arXiv:1903.11430}.

\bibitem[{\citenamefont{G$\acute{\textrm{e}}$rard and
  Kou}(2005)}]{Gerard:2004gx}
\bibinfo{author}{\bibfnamefont{J.~M.} \bibnamefont{G$\acute{\textrm{e}}$rard}}
  \bibnamefont{and} \bibinfo{author}{\bibfnamefont{E.}~\bibnamefont{Kou}},
  \bibinfo{journal}{Phys. Lett.} \textbf{\bibinfo{volume}{B616}},
  \bibinfo{pages}{85} (\bibinfo{year}{2005}), \eprint{arXiv:hep-ph/0411292}.

\bibitem[{\citenamefont{Zhao}(2011)}]{Zhao:2010mm}
\bibinfo{author}{\bibfnamefont{Q.}~\bibnamefont{Zhao}}, \bibinfo{journal}{Phys.
  Lett.} \textbf{\bibinfo{volume}{B697}}, \bibinfo{pages}{52}
  (\bibinfo{year}{2011}), \eprint{arXiv:1012.1165}.

\bibitem[{\citenamefont{G$\acute{\textrm{e}}$rard and
  Martini}(2014)}]{Gerard:2013gya}
\bibinfo{author}{\bibfnamefont{J.-M.} \bibnamefont{G$\acute{\textrm{e}}$rard}}
  \bibnamefont{and} \bibinfo{author}{\bibfnamefont{A.}~\bibnamefont{Martini}},
  \bibinfo{journal}{Phys. Lett.} \textbf{\bibinfo{volume}{B730}},
  \bibinfo{pages}{264} (\bibinfo{year}{2014}), \eprint{arXiv:1312.3081}.

\bibitem[{\citenamefont{Escribano et~al.}(2014)\citenamefont{Escribano,
  Masjuan, and Sanchez-Puertas}}]{Escribano:2013kba}
\bibinfo{author}{\bibfnamefont{R.}~\bibnamefont{Escribano}},
  \bibinfo{author}{\bibfnamefont{P.}~\bibnamefont{Masjuan}}, \bibnamefont{and}
  \bibinfo{author}{\bibfnamefont{P.}~\bibnamefont{Sanchez-Puertas}},
  \bibinfo{journal}{Phys. Rev.} \textbf{\bibinfo{volume}{D89}},
  \bibinfo{pages}{034014} (\bibinfo{year}{2014}), \eprint{arXiv:1307.2061}.

\bibitem[{\citenamefont{Feldmann et~al.}(1998)\citenamefont{Feldmann, Kroll,
  and Stech}}]{Feldmann:1998vh}
\bibinfo{author}{\bibfnamefont{T.}~\bibnamefont{Feldmann}},
  \bibinfo{author}{\bibfnamefont{P.}~\bibnamefont{Kroll}}, \bibnamefont{and}
  \bibinfo{author}{\bibfnamefont{B.}~\bibnamefont{Stech}},
  \bibinfo{journal}{Phys. Rev.} \textbf{\bibinfo{volume}{D58}},
  \bibinfo{pages}{114006} (\bibinfo{year}{1998}), \eprint{arXiv:hep-ph/9802409}.

\bibitem[{\citenamefont{Ablikim et~al.}(2016)}]{Ablikim:2016uoc}
\bibinfo{author}{\bibfnamefont{M.}~\bibnamefont{Ablikim}} \bibnamefont{et~al.}
  (\bibinfo{collaboration}{BESIII}), \bibinfo{journal}{Phys. Rev. Lett.}
  \textbf{\bibinfo{volume}{116}}, \bibinfo{pages}{251802}
  (\bibinfo{year}{2016}), \eprint{arXiv:1603.04936}.

\bibitem[{\citenamefont{Tanabashi et~al.}(2018)}]{Tanabashi:2018oca}
\bibinfo{author}{\bibfnamefont{M.}~\bibnamefont{Tanabashi}}
  \bibnamefont{et~al.} (\bibinfo{collaboration}{Particle Data Group}),
  \bibinfo{journal}{Phys. Rev.} \textbf{\bibinfo{volume}{D98}},
  \bibinfo{pages}{030001} (\bibinfo{year}{2018}).

\bibitem[{\citenamefont{Zhu and Dai}(2016)}]{Zhu:2016udl}
\bibinfo{author}{\bibfnamefont{R.}~\bibnamefont{Zhu}} \bibnamefont{and}
  \bibinfo{author}{\bibfnamefont{J.-P.} \bibnamefont{Dai}},
  \bibinfo{journal}{Phys. Rev.} \textbf{\bibinfo{volume}{D94}},
  \bibinfo{pages}{094034} (\bibinfo{year}{2016}), \eprint{arXiv:1610.00288}.

\bibitem[{\citenamefont{Wu et~al.}(2017)\citenamefont{Wu, Li, and
  Zhang}}]{Wu:2017pep}
\bibinfo{author}{\bibfnamefont{Q.}~\bibnamefont{Wu}},
  \bibinfo{author}{\bibfnamefont{G.}~\bibnamefont{Li}}, \bibnamefont{and}
  \bibinfo{author}{\bibfnamefont{Y.}~\bibnamefont{Zhang}},
  \bibinfo{journal}{Eur. Phys. J.} \textbf{\bibinfo{volume}{C77}},
  \bibinfo{pages}{336} (\bibinfo{year}{2017}), \eprint{arXiv:1705.04409}.

\bibitem[{\citenamefont{Buchmuller and Tye}(1981)}]{Buchmuller:1980su}
\bibinfo{author}{\bibfnamefont{W.}~\bibnamefont{Buchmuller}} \bibnamefont{and}
  \bibinfo{author}{\bibfnamefont{S.~H.~H.} \bibnamefont{Tye}},
  \bibinfo{journal}{Phys. Rev.} \textbf{\bibinfo{volume}{D24}},
  \bibinfo{pages}{132} (\bibinfo{year}{1981}).

\bibitem[{\citenamefont{Eichten and Quigg}(1995)}]{Eichten:1995ch}
\bibinfo{author}{\bibfnamefont{E.~J.} \bibnamefont{Eichten}} \bibnamefont{and}
  \bibinfo{author}{\bibfnamefont{C.}~\bibnamefont{Quigg}},
  \bibinfo{journal}{Phys. Rev.} \textbf{\bibinfo{volume}{D52}},
  \bibinfo{pages}{1726} (\bibinfo{year}{1995}), \eprint{arXiv:hep-ph/9503356}.

\bibitem[{\citenamefont{Bodwin et~al.}(1995)\citenamefont{Bodwin, Braaten, and
  Lepage}}]{Bodwin:1994jh}
\bibinfo{author}{\bibfnamefont{G.~T.} \bibnamefont{Bodwin}},
  \bibinfo{author}{\bibfnamefont{E.}~\bibnamefont{Braaten}}, \bibnamefont{and}
  \bibinfo{author}{\bibfnamefont{G.~P.} \bibnamefont{Lepage}},
  \bibinfo{journal}{Phys. Rev.} \textbf{\bibinfo{volume}{D51}},
  \bibinfo{pages}{1125} (\bibinfo{year}{1995}), \bibinfo{note}{[Erratum: Phys.
  Rev.D55,5853(1997)]}, \eprint{arXiv:hep-ph/9407339}.

\bibitem[{\citenamefont{Ali and Parkhomenko}(2002)}]{Ali:2000ci}
\bibinfo{author}{\bibfnamefont{A.}~\bibnamefont{Ali}} \bibnamefont{and}
  \bibinfo{author}{\bibfnamefont{{\relax Ya}.}~\bibnamefont{Parkhomenko}},
  \bibinfo{journal}{Phys. Rev.} \textbf{\bibinfo{volume}{D65}},
  \bibinfo{pages}{074020} (\bibinfo{year}{2002}), \eprint{arXiv:hep-ph/0012212}.

\bibitem[{\citenamefont{Appelquist and Politzer}(1975)}]{Appelquist:1974zd}
\bibinfo{author}{\bibfnamefont{T.}~\bibnamefont{Appelquist}} \bibnamefont{and}
  \bibinfo{author}{\bibfnamefont{H.~D.} \bibnamefont{Politzer}},
  \bibinfo{journal}{Phys. Rev. Lett.} \textbf{\bibinfo{volume}{34}},
  \bibinfo{pages}{43} (\bibinfo{year}{1975}).

\bibitem[{\citenamefont{De~R$\acute{\textrm{u}}$jula and
  Glashow}(1975)}]{DeRujula:1974rkb}
\bibinfo{author}{\bibfnamefont{A.}~\bibnamefont{De~R$\acute{\textrm{u}}$jula}}
  \bibnamefont{and} \bibinfo{author}{\bibfnamefont{S.~L.}
  \bibnamefont{Glashow}}, \bibinfo{journal}{Phys. Rev. Lett.}
  \textbf{\bibinfo{volume}{34}}, \bibinfo{pages}{46} (\bibinfo{year}{1975}).

\bibitem[{\citenamefont{Barbieri
  et~al.}(1976{\natexlab{a}})\citenamefont{Barbieri, Gatto, and
  K$\ddot{\textrm{o}}$gerler}}]{Barbieri:1975am}
\bibinfo{author}{\bibfnamefont{R.}~\bibnamefont{Barbieri}},
  \bibinfo{author}{\bibfnamefont{R.}~\bibnamefont{Gatto}}, \bibnamefont{and}
  \bibinfo{author}{\bibfnamefont{R.}~\bibnamefont{K$\ddot{\textrm{o}}$gerler}},
  \bibinfo{journal}{Phys. Lett.} \textbf{\bibinfo{volume}{B60}},
  \bibinfo{pages}{183} (\bibinfo{year}{1976}{\natexlab{a}}).

\bibitem[{\citenamefont{Novikov et~al.}(1978)\citenamefont{Novikov, Okun,
  Shifman, Vainshtein, Voloshin, and Zakharov}}]{Novikov:1977dq}
\bibinfo{author}{\bibfnamefont{V.~A.} \bibnamefont{Novikov}},
  \bibinfo{author}{\bibfnamefont{L.~B.} \bibnamefont{Okun}},
  \bibinfo{author}{\bibfnamefont{M.~A.} \bibnamefont{Shifman}},
  \bibinfo{author}{\bibfnamefont{A.~I.} \bibnamefont{Vainshtein}},
  \bibinfo{author}{\bibfnamefont{M.~B.} \bibnamefont{Voloshin}},
  \bibnamefont{and} \bibinfo{author}{\bibfnamefont{V.~I.}
  \bibnamefont{Zakharov}}, \bibinfo{journal}{Phys. Rept.}
  \textbf{\bibinfo{volume}{41}}, \bibinfo{pages}{1} (\bibinfo{year}{1978}).

\bibitem[{\citenamefont{Barbieri et~al.}(1979)\citenamefont{Barbieri, d'Emilio,
  Curci, and Remiddi}}]{Barbieri:1979be}
\bibinfo{author}{\bibfnamefont{R.}~\bibnamefont{Barbieri}},
  \bibinfo{author}{\bibfnamefont{E.}~\bibnamefont{d'Emilio}},
  \bibinfo{author}{\bibfnamefont{G.}~\bibnamefont{Curci}}, \bibnamefont{and}
  \bibinfo{author}{\bibfnamefont{E.}~\bibnamefont{Remiddi}},
  \bibinfo{journal}{Nucl. Phys.} \textbf{\bibinfo{volume}{B154}},
  \bibinfo{pages}{535} (\bibinfo{year}{1979}).

\bibitem[{\citenamefont{Mackenzie and Lepage}(1981)}]{Mackenzie:1981sf}
\bibinfo{author}{\bibfnamefont{P.~B.} \bibnamefont{Mackenzie}}
  \bibnamefont{and} \bibinfo{author}{\bibfnamefont{G.~P.}
  \bibnamefont{Lepage}}, \bibinfo{journal}{Phys. Rev. Lett.}
  \textbf{\bibinfo{volume}{47}}, \bibinfo{pages}{1244} (\bibinfo{year}{1981}).

\bibitem[{\citenamefont{K$\ddot{\textrm{o}}$rner
  et~al.}(1983)\citenamefont{K$\ddot{\textrm{o}}$rner, K$\ddot{\textrm{u}}$hn,
  Krammer, and Schneider}}]{Korner:1982vg}
\bibinfo{author}{\bibfnamefont{J.~G.} \bibnamefont{K$\ddot{\textrm{o}}$rner}},
  \bibinfo{author}{\bibfnamefont{J.~H.} \bibnamefont{K$\ddot{\textrm{u}}$hn}},
  \bibinfo{author}{\bibfnamefont{M.}~\bibnamefont{Krammer}}, \bibnamefont{and}
  \bibinfo{author}{\bibfnamefont{H.}~\bibnamefont{Schneider}},
  \bibinfo{journal}{Nucl. Phys.} \textbf{\bibinfo{volume}{B229}},
  \bibinfo{pages}{115} (\bibinfo{year}{1983}).

\bibitem[{\citenamefont{K$\ddot{\textrm{u}}$hn}(1983)}]{Kuhn:1983yr}
\bibinfo{author}{\bibfnamefont{J.~H.} \bibnamefont{K$\ddot{\textrm{u}}$hn}},
  \bibinfo{journal}{Phys. Lett.} \textbf{\bibinfo{volume}{B127}},
  \bibinfo{pages}{257} (\bibinfo{year}{1983}).

\bibitem[{\citenamefont{Kroll and
  Passek-Kumeri$\check{\textrm{c}}$ki}(2003)}]{Kroll:2002nt}
\bibinfo{author}{\bibfnamefont{P.}~\bibnamefont{Kroll}} \bibnamefont{and}
  \bibinfo{author}{\bibfnamefont{K.}~\bibnamefont{Passek-Kumeri$\check{\textrm{c}}$ki}},
  \bibinfo{journal}{Phys. Rev.} \textbf{\bibinfo{volume}{D67}},
  \bibinfo{pages}{054017} (\bibinfo{year}{2003}), \eprint{arXiv:hep-ph/0210045}.

\bibitem[{\citenamefont{Baier and Grozin}(1981)}]{Baier:1981pm}
\bibinfo{author}{\bibfnamefont{V.~N.} \bibnamefont{Baier}} \bibnamefont{and}
  \bibinfo{author}{\bibfnamefont{A.~G.} \bibnamefont{Grozin}},
  \bibinfo{journal}{Nucl. Phys.} \textbf{\bibinfo{volume}{B192}},
  \bibinfo{pages}{476} (\bibinfo{year}{1981}).

\bibitem[{\citenamefont{Guberina and
  K$\ddot{\textrm{u}}$hn}(1981)}]{Guberina:1980xb}
\bibinfo{author}{\bibfnamefont{B.}~\bibnamefont{Guberina}} \bibnamefont{and}
  \bibinfo{author}{\bibfnamefont{J.~H.} \bibnamefont{K$\ddot{\textrm{u}}$hn}},
  \bibinfo{journal}{Lett. Nuovo Cim.} \textbf{\bibinfo{volume}{32}},
  \bibinfo{pages}{295} (\bibinfo{year}{1981}).

\bibitem[{\citenamefont{K$\ddot{\textrm{u}}$hn
  et~al.}(1979)\citenamefont{K$\ddot{\textrm{u}}$hn, Kaplan, and
  Safiani}}]{Kuhn:1979bb}
\bibinfo{author}{\bibfnamefont{J.~H.} \bibnamefont{K$\ddot{\textrm{u}}$hn}},
  \bibinfo{author}{\bibfnamefont{J.}~\bibnamefont{Kaplan}}, \bibnamefont{and}
  \bibinfo{author}{\bibfnamefont{E.~G.~O.} \bibnamefont{Safiani}},
  \bibinfo{journal}{Nucl. Phys.} \textbf{\bibinfo{volume}{B157}},
  \bibinfo{pages}{125} (\bibinfo{year}{1979}).

\bibitem[{\citenamefont{Guberina et~al.}(1980)\citenamefont{Guberina,
  K$\ddot{\textrm{u}}$hn, Peccei, and
  R$\ddot{\textrm{u}}$ckl}}]{Guberina:1980dc}
\bibinfo{author}{\bibfnamefont{B.}~\bibnamefont{Guberina}},
  \bibinfo{author}{\bibfnamefont{J.~H.} \bibnamefont{K$\ddot{\textrm{u}}$hn}},
  \bibinfo{author}{\bibfnamefont{R.~D.} \bibnamefont{Peccei}},
  \bibnamefont{and}
  \bibinfo{author}{\bibfnamefont{R.}~\bibnamefont{R$\ddot{\textrm{u}}$ckl}},
  \bibinfo{journal}{Nucl. Phys.} \textbf{\bibinfo{volume}{B174}},
  \bibinfo{pages}{317} (\bibinfo{year}{1980}).

\bibitem[{\citenamefont{Ball and Jones}(2007)}]{Ball:2007hb}
\bibinfo{author}{\bibfnamefont{P.}~\bibnamefont{Ball}} \bibnamefont{and}
  \bibinfo{author}{\bibfnamefont{G.~W.} \bibnamefont{Jones}},
  \bibinfo{journal}{JHEP} \textbf{\bibinfo{volume}{08}}, \bibinfo{pages}{025}
  (\bibinfo{year}{2007}), \eprint{arXiv:0706.3628}.

\bibitem[{\citenamefont{Chernyak and Zhitnitsky}(1984)}]{Chernyak:1983ej}
\bibinfo{author}{\bibfnamefont{V.~L.} \bibnamefont{Chernyak}} \bibnamefont{and}
  \bibinfo{author}{\bibfnamefont{A.~R.} \bibnamefont{Zhitnitsky}},
  \bibinfo{journal}{Phys. Rept.} \textbf{\bibinfo{volume}{112}},
  \bibinfo{pages}{173} (\bibinfo{year}{1984}).

\bibitem[{\citenamefont{Muta and Yang}(2000)}]{Muta:1999tc}
\bibinfo{author}{\bibfnamefont{T.}~\bibnamefont{Muta}} \bibnamefont{and}
  \bibinfo{author}{\bibfnamefont{M.-Z.} \bibnamefont{Yang}},
  \bibinfo{journal}{Phys. Rev.} \textbf{\bibinfo{volume}{D61}},
  \bibinfo{pages}{054007} (\bibinfo{year}{2000}), \eprint{arXiv:hep-ph/9909484}.

\bibitem[{\citenamefont{Yang and Yang}(2001)}]{Yang:2000ce}
\bibinfo{author}{\bibfnamefont{M.-Z.} \bibnamefont{Yang}} \bibnamefont{and}
  \bibinfo{author}{\bibfnamefont{Y.-D.} \bibnamefont{Yang}},
  \bibinfo{journal}{Nucl. Phys.} \textbf{\bibinfo{volume}{B609}},
  \bibinfo{pages}{469} (\bibinfo{year}{2001}), \eprint{arXiv:hep-ph/0012208}.

\bibitem[{\citenamefont{Agaev et~al.}(2014)\citenamefont{Agaev, Braun, Offen,
  Porkert, and Sch$\ddot{\textrm{a}}$fer}}]{Agaev:2014wna}
\bibinfo{author}{\bibfnamefont{S.~S.} \bibnamefont{Agaev}},
  \bibinfo{author}{\bibfnamefont{V.~M.} \bibnamefont{Braun}},
  \bibinfo{author}{\bibfnamefont{N.}~\bibnamefont{Offen}},
  \bibinfo{author}{\bibfnamefont{F.~A.} \bibnamefont{Porkert}},
  \bibnamefont{and}
  \bibinfo{author}{\bibfnamefont{A.}~\bibnamefont{Sch$\ddot{\textrm{a}}$fer}},
  \bibinfo{journal}{Phys. Rev.} \textbf{\bibinfo{volume}{D90}},
  \bibinfo{pages}{074019} (\bibinfo{year}{2014}), \eprint{arXiv:1409.4311}.

\bibitem[{\citenamefont{Dittmaier}(2003)}]{Dittmaier:2003bc}
\bibinfo{author}{\bibfnamefont{S.}~\bibnamefont{Dittmaier}},
  \bibinfo{journal}{Nucl. Phys.} \textbf{\bibinfo{volume}{B675}},
  \bibinfo{pages}{447} (\bibinfo{year}{2003}), \eprint{arXiv:hep-ph/0308246}.

\bibitem[{\citenamefont{'t~Hooft and Veltman}(1979)}]{tHooft:1978jhc}
\bibinfo{author}{\bibfnamefont{G.}~\bibnamefont{'t~Hooft}} \bibnamefont{and}
  \bibinfo{author}{\bibfnamefont{M.~J.~G.} \bibnamefont{Veltman}},
  \bibinfo{journal}{Nucl. Phys.} \textbf{\bibinfo{volume}{B153}},
  \bibinfo{pages}{365} (\bibinfo{year}{1979}).

\bibitem[{\citenamefont{Patel}(2015)}]{Patel:2015tea}
\bibinfo{author}{\bibfnamefont{H.~H.} \bibnamefont{Patel}},
  \bibinfo{journal}{Comput. Phys. Commun.} \textbf{\bibinfo{volume}{197}},
  \bibinfo{pages}{276} (\bibinfo{year}{2015}), \eprint{arXiv:1503.01469}.

\bibitem[{\citenamefont{Patel}(2017)}]{Patel:2016fam}
\bibinfo{author}{\bibfnamefont{H.~H.} \bibnamefont{Patel}},
  \bibinfo{journal}{Comput. Phys. Commun.} \textbf{\bibinfo{volume}{218}},
  \bibinfo{pages}{66} (\bibinfo{year}{2017}), \eprint{arXiv:1612.00009}.

\bibitem[{\citenamefont{Alte et~al.}(2016)\citenamefont{Alte,
  K$\ddot{\textrm{o}}$nig, and Neubert}}]{Alte:2015dpo}
\bibinfo{author}{\bibfnamefont{S.}~\bibnamefont{Alte}},
  \bibinfo{author}{\bibfnamefont{M.}~\bibnamefont{K$\ddot{\textrm{o}}$nig}},
  \bibnamefont{and} \bibinfo{author}{\bibfnamefont{M.}~\bibnamefont{Neubert}},
  \bibinfo{journal}{JHEP} \textbf{\bibinfo{volume}{02}}, \bibinfo{pages}{162}
  (\bibinfo{year}{2016}), \eprint{arXiv:1512.09135}.

\bibitem[{\citenamefont{Eichten et~al.}(1978)\citenamefont{Eichten, Gottfried,
  Kinoshita, Lane, and Yan}}]{Eichten:1978tg}
\bibinfo{author}{\bibfnamefont{E.}~\bibnamefont{Eichten}},
  \bibinfo{author}{\bibfnamefont{K.}~\bibnamefont{Gottfried}},
  \bibinfo{author}{\bibfnamefont{T.}~\bibnamefont{Kinoshita}},
  \bibinfo{author}{\bibfnamefont{K.~D.} \bibnamefont{Lane}}, \bibnamefont{and}
  \bibinfo{author}{\bibfnamefont{T.-M.} \bibnamefont{Yan}},
  \bibinfo{journal}{Phys. Rev.} \textbf{\bibinfo{volume}{D17}},
  \bibinfo{pages}{3090} (\bibinfo{year}{1978}), \bibinfo{note}{[Erratum: Phys.
  Rev.D21,313(1980)]}.

\bibitem[{\citenamefont{Eichten et~al.}(1980)\citenamefont{Eichten, Gottfried,
  Kinoshita, Lane, and Yan}}]{Eichten:1979ms}
\bibinfo{author}{\bibfnamefont{E.}~\bibnamefont{Eichten}},
  \bibinfo{author}{\bibfnamefont{K.}~\bibnamefont{Gottfried}},
  \bibinfo{author}{\bibfnamefont{T.}~\bibnamefont{Kinoshita}},
  \bibinfo{author}{\bibfnamefont{K.~D.} \bibnamefont{Lane}}, \bibnamefont{and}
  \bibinfo{author}{\bibfnamefont{T.-M.} \bibnamefont{Yan}},
  \bibinfo{journal}{Phys. Rev.} \textbf{\bibinfo{volume}{D21}},
  \bibinfo{pages}{203} (\bibinfo{year}{1980}).

\bibitem[{\citenamefont{Feldmann et~al.}(1999)\citenamefont{Feldmann, Kroll,
  and Stech}}]{Feldmann:1998sh}
\bibinfo{author}{\bibfnamefont{T.}~\bibnamefont{Feldmann}},
  \bibinfo{author}{\bibfnamefont{P.}~\bibnamefont{Kroll}}, \bibnamefont{and}
  \bibinfo{author}{\bibfnamefont{B.}~\bibnamefont{Stech}},
  \bibinfo{journal}{Phys. Lett.} \textbf{\bibinfo{volume}{B449}},
  \bibinfo{pages}{339} (\bibinfo{year}{1999}), \eprint{arXiv:hep-ph/9812269}.

\bibitem[{\citenamefont{Feldmann}(2000)}]{Feldmann:1999uf}
\bibinfo{author}{\bibfnamefont{T.}~\bibnamefont{Feldmann}},
  \bibinfo{journal}{Int. J. Mod. Phys.} \textbf{\bibinfo{volume}{A15}},
  \bibinfo{pages}{159} (\bibinfo{year}{2000}), \eprint{arXiv:hep-ph/9907491}.

\bibitem[{\citenamefont{Escribano and
  Fr$\grave{\textrm{e}}$re}(2005)}]{Escribano:2005qq}
\bibinfo{author}{\bibfnamefont{R.}~\bibnamefont{Escribano}} \bibnamefont{and}
  \bibinfo{author}{\bibfnamefont{J.-M.}
  \bibnamefont{Fr$\grave{\textrm{e}}$re}}, \bibinfo{journal}{JHEP}
  \textbf{\bibinfo{volume}{06}}, \bibinfo{pages}{029} (\bibinfo{year}{2005}),
  \eprint{arXiv:hep-ph/0501072}.

\bibitem[{\citenamefont{Escribano and Nadal}(2007)}]{Escribano:2007cd}
\bibinfo{author}{\bibfnamefont{R.}~\bibnamefont{Escribano}} \bibnamefont{and}
  \bibinfo{author}{\bibfnamefont{J.}~\bibnamefont{Nadal}},
  \bibinfo{journal}{JHEP} \textbf{\bibinfo{volume}{05}}, \bibinfo{pages}{006}
  (\bibinfo{year}{2007}), \eprint{arXiv:hep-ph/0703187}.

\bibitem[{\citenamefont{Cao}(2012)}]{Cao:2012nj}
\bibinfo{author}{\bibfnamefont{F.-G.} \bibnamefont{Cao}},
  \bibinfo{journal}{Phys. Rev.} \textbf{\bibinfo{volume}{D85}},
  \bibinfo{pages}{057501} (\bibinfo{year}{2012}), \eprint{arXiv:1202.6075}.

\bibitem[{\citenamefont{Aubert et~al.}(2006)}]{Aubert:2006cy}
\bibinfo{author}{\bibfnamefont{B.}~\bibnamefont{Aubert}} \bibnamefont{et~al.}
  (\bibinfo{collaboration}{BaBar}), \bibinfo{journal}{Phys. Rev.}
  \textbf{\bibinfo{volume}{D74}}, \bibinfo{pages}{012002}
  (\bibinfo{year}{2006}), \eprint{arXiv:hep-ex/0605018}.

\bibitem[{\citenamefont{Ablikim et~al.}(2012)}]{Ablikim:2012ur}
\bibinfo{author}{\bibfnamefont{M.}~\bibnamefont{Ablikim}} \bibnamefont{et~al.}
  (\bibinfo{collaboration}{BESIII}), \bibinfo{journal}{Phys. Rev.}
  \textbf{\bibinfo{volume}{D86}}, \bibinfo{pages}{092009}
  (\bibinfo{year}{2012}), \eprint{arXiv:1209.4963}.

\bibitem[{\citenamefont{Babusci et~al.}(2013)}]{Babusci:2012ik}
\bibinfo{author}{\bibfnamefont{D.}~\bibnamefont{Babusci}} \bibnamefont{et~al.}
  (\bibinfo{collaboration}{KLOE-2}), \bibinfo{journal}{JHEP}
  \textbf{\bibinfo{volume}{01}}, \bibinfo{pages}{119} (\bibinfo{year}{2013}),
  \eprint{arXiv:1211.1845}.

\bibitem[{\citenamefont{Gregory et~al.}(2012)\citenamefont{Gregory, Irving,
  Richards, and McNeile}}]{Gregory:2011sg}
\bibinfo{author}{\bibfnamefont{E.~B.} \bibnamefont{Gregory}},
  \bibinfo{author}{\bibfnamefont{A.~C.} \bibnamefont{Irving}},
  \bibinfo{author}{\bibfnamefont{C.~M.} \bibnamefont{Richards}},
  \bibnamefont{and} \bibinfo{author}{\bibfnamefont{C.}~\bibnamefont{McNeile}}
  (\bibinfo{collaboration}{UKQCD}), \bibinfo{journal}{Phys. Rev.}
  \textbf{\bibinfo{volume}{D86}}, \bibinfo{pages}{014504}
  (\bibinfo{year}{2012}), \eprint{arXiv:1112.4384}.

\bibitem[{\citenamefont{Ottnad and Urbach}(2018)}]{Ottnad:2017bjt}
\bibinfo{author}{\bibfnamefont{K.}~\bibnamefont{Ottnad}} \bibnamefont{and}
  \bibinfo{author}{\bibfnamefont{C.}~\bibnamefont{Urbach}}
  (\bibinfo{collaboration}{ETM}), \bibinfo{journal}{Phys. Rev.}
  \textbf{\bibinfo{volume}{D97}}, \bibinfo{pages}{054508}
  (\bibinfo{year}{2018}), \eprint{arXiv:1710.07986}.

\bibitem[{\citenamefont{Barbieri
  et~al.}(1976{\natexlab{b}})\citenamefont{Barbieri, Gatto, and
  Remiddi}}]{Barbieri:1976fp}
\bibinfo{author}{\bibfnamefont{R.}~\bibnamefont{Barbieri}},
  \bibinfo{author}{\bibfnamefont{R.}~\bibnamefont{Gatto}}, \bibnamefont{and}
  \bibinfo{author}{\bibfnamefont{E.}~\bibnamefont{Remiddi}},
  \bibinfo{journal}{Phys. Lett.} \textbf{\bibinfo{volume}{B61}},
  \bibinfo{pages}{465} (\bibinfo{year}{1976}{\natexlab{b}}).

\bibitem[{\citenamefont{Bodwin et~al.}(1992)\citenamefont{Bodwin, Braaten, and
  Lepage}}]{Bodwin:1992ye}
\bibinfo{author}{\bibfnamefont{G.~T.} \bibnamefont{Bodwin}},
  \bibinfo{author}{\bibfnamefont{E.}~\bibnamefont{Braaten}}, \bibnamefont{and}
  \bibinfo{author}{\bibfnamefont{G.~P.} \bibnamefont{Lepage}},
  \bibinfo{journal}{Phys. Rev.} \textbf{\bibinfo{volume}{D46}},
  \bibinfo{pages}{R1914} (\bibinfo{year}{1992}), \eprint{arXiv:hep-lat/9205006}.

\bibitem[{\citenamefont{Kroll}(1998)}]{Kroll:1997vt}
\bibinfo{author}{\bibfnamefont{P.}~\bibnamefont{Kroll}},
  \bibinfo{journal}{Nucl. Phys. Proc. Suppl.} \textbf{\bibinfo{volume}{64}},
  \bibinfo{pages}{456} (\bibinfo{year}{1998}), \eprint{arXiv:hep-ph/9709393}.

\bibitem[{\citenamefont{Wong}(1999)}]{Wong:1998rv}
\bibinfo{author}{\bibfnamefont{S.~M.~H.} \bibnamefont{Wong}},
  \bibinfo{journal}{Nucl. Phys. Proc. Suppl.} \textbf{\bibinfo{volume}{74}},
  \bibinfo{pages}{231} (\bibinfo{year}{1999}), \eprint{arXiv:hep-ph/9809447}.

\end{thebibliography}

\end{CJK*}
\end{document}